\def \vc #1{{\mbox{\boldmath $#1$}}}

\def\boldk{{\bf k}}
\def\boldx{{\bf x}}
\def\boldg{{\bf g}}

\def\thetag{{\vc \theta}}

\def\alphag{{\vc \alpha}}
\def\deltag{{\vc \delta}}
\def\gammag{{\vc \gamma}}
\def\epsilong{{\vc \epsilon}}
\def\taug{{\vc \tau}}
\documentclass[11pt]{article}
\usepackage{psfig}
\begin{document}
\noindent {\Large PROBING THE UNIVERSE WITH WEAK LENSING}
\vskip 5truemm
\noindent {\large {\it Yannick Mellier$^{1,2}$}}
\vskip 5truemm
$^1${\small Institut d'Astrophysique de Paris, 98 bis Boulevard Arago,
75014 Paris, France.}\\
$^2${\small Observatoire de Paris, DEMIRM, 61 avenue de l'Observatoire, 
75014 Paris, France.\\
E-mail: mellier@iap.fr}
\vskip 5truemm
\centerline{ABSTRACT} 

Gravitational lenses can provide crucial information on the geometry of the Universe,  
 on the cosmological scenario of formation of its structures as well as on the history 
 of its components with look-back time. In this review, I focus on the most recent 
results obtained during the last five years from the analysis of the 
weak lensing regime. The interest of weak lensing as a probe of dark matter  and the 
 for study of the coupling between light and mass  on scales of  clusters of 
galaxies, large scale structures and  galaxies is discussed first.  Then I 
 present the impact of weak lensing for the study of distant galaxies and of 
 the 
 population of lensed sources as function of redshift. Finally, I discuss the potential 
interest of weak lensing to constrain the cosmological parameters, either from 
 pure geometrical effects observed in peculiar lenses, or from the coupling of 
 weak lensing  with the CMB. \\ 
 
\vskip 5truemm
To appear in Vol.~37 of {\it Annual Review of Astronomy and Astrophysics}
\vskip 5truemm
\hfill
\newpage
\tableofcontents
\newpage
\section{INTRODUCTION}
Matter intervening along the light pathes of photons causes
a displacement and a distortion of ray bundles. The properties and the
interpretation of this effect depend on   
 the projected mass density integrated along the line of sight  
 and on  the cosmological angular distances to the observer,
the lens and the source.  \\
The sensitivity to  mass density implies  
that gravitational lensing effects can probe the mass of deflectors,
regardless their dynamical stage and the nature of the deflecting matter.
This is therefore a unique tool to probe the dark matter distribution 
in gravitational systems as well as to study the 
 dynamical  evolution of structures with redshift.  The
dependence on the various angular distances involved in the 
lens configuration means that the deviation
angle depends on the cosmological parameters, $H_o$, $\Omega$ and
$\lambda$, so that the analysis of gravitational lensing can potentially 
 provide a diagnosis on cosmography.  Of course, the 
sensitivity to cosmological parameters is
not unique to gravitational lensing 
 and many other astrophysical phenomena depend on them.
 However, due to magnification, image multiplicity and
deflection angle produced by lensing, it is possible to use the lensing
 effect   
 as a bonus when compared to other experiments: image
  magnification permits to observe the high-redshift universe, 
 to study evolution of galaxies with look-back time and to 
 compare with theoretical cosmological scenarios. Image multiplicity 
 probes different light paths taken by photons emitted by one source. 
 By computing time delays of the same transient event observed in each
individual image, one can measure $H_o$.
Finally, for high-redshift sources the deflecting angle depends on the 
 geometry of the universe  and provides a unique tool 
to measure  the cosmological parameters. \\ 
The interest of gravitational lensing for cosmology started 
 very early, after  Zwicky's discovery  (Zwicky 1933) of the apparent contradiction
between the visible mass of the  Coma cluster and its virial mass,
 which could not be explained without invoking that it is   
dominated by {\it unseen mass}.  This surprising statement  could not be confirmed 
 without an independent mass estimator, which could probe the total mass directly,  
 without using the light distribution or  critical assumptions on the 
 dynamical stage of the cluster components. Four years latter,  Zwicky (1937)  envisioned 
that {\it extragalactic nebulae} could be efficient gravitational lenses and 
 provide an invaluable tool for weighting the gravitational systems of the Universe.  \\
The other  works which raised the cosmological 
interest of lensing for cosmology are more contemporary.  Refsdal (1964)
first emphasized that time delays in multiple images could be used 
 to measure $H_o$, and the very first considerations about light 
propagation and 
deformation of ray bundles in inhomogeneous universes were discussed 
 initially by Sachs (1961), Zel'dovich (1964) and later by Gunn (1967). 
 From the observational point of view, the discovery of the first 
multiply imaged quasar (Walsh, Carswell \& Weymann 1979) and the 
 first distorted galaxies (Soucail et al  1987, Lynds \& Petrosian 1986)
 were major steps which boosted theoretical and observational
investigations  of gravitational lenses. \\
Most of the cosmological interest of gravitational lenses has already
been reviewed in Blandford \& Narayan (1992), Schneider, Ehlers \& Falco
(1992) and Refsdal \& Surdej (1994).  Fort \& Mellier (1994) presented the first review 
which  focussed 
particularly on the use of arc(let)s  in cosmology,  
 and the interest of lensed galaxies 
 to probe the deep universe has been also recently reviewed by Ellis (1997).  With the amazing 
 observational and theoretical developments in the field, in particular 
 in weak lensing, it seems timely to review all these results and to address
the new and future issues in the area.  \\ 
During the last five years, thanks to
the seminal work on mass reconstruction from weak lensing analyses
(Tyson, Valdes \& Wenk 1990, and Kaiser \& Squires 1993), the works devoted to mass 
reconstruction 
algorithms have provided new and robust tools to study the mass
distribution of gravitational systems and have permitted to establish a link between 
 theoretical investigations of weak lensing and the observations 
 of weakly distorted galaxies. In particular, impressive developments 
 have been done on the cosmological diagnoses  
 from the analysis of weak lensing by large scale structures. 
  Theoretical and numerical studies demonstrate that the statistical 
 analysis of gravitational lensing will provide valuable insights 
 on the mass distribution as well as on the cosmological parameters. With
the coming of new wide field surveys with subarcsecond seeing (like 
 MEGACAM at the Canada-France-Hawaii Telescope (CFHT) or the VLT-Survey-Telescope (VST) at Paranal) or very wide field shallow surveys  
(like the VLA-FIRST survey or the SLOAN Digital Sky Survey (SDSS)), weak lensing 
analysis should 
 probe the power spectrum of the projected mass density,
   from arcminutes up to degree scales.  Visible   
 weak lensing surveys should also be capable to provide a projected
mass map of the universe, exactly like the APM survey provides the
visible light distribution (Maddox et al 1990). From the observational point of view, the
outstanding images coming from Hubble Space Telescope (HST) had a 
considerable positive impact on our 
 intuitions about the potential usefulness of gravitational distortion.
The wonderful shear pattern around the lensing cluster A2218 is a 
visual proof in itself that weak lensing works and that it reveals 
 directly the mass distribution.  One of the most spectacular use of
 HST images for lensing was done by Kneib et al  (1996b), also  in A2218.  
 The superb HST images allowed them to demonstrate, only from the
morphology of one arclet, and without the need of 
  a spectroscopic redshift, that it must be  
 a lensed image associated to the same source as the giant arc.  
The similarity of the morphologies of the
giant arc and the counter-image is so impressive that it cannot be 
 questioned that they are images of the same source. In parallel, the Keck telescope 
 which is currently detecting the most distant galaxies, reveals the obvious   
 interest of {\it giant gravitational  telescopes}. 
 Finally, the impressive results obtained by SCUBA in the submillimeter
wavebands have shown that the joint use of a submillimeter instrument with 
 magnification of high-redshift galaxies is an ideal tool to study 
 the evolution and content of distant galaxies.
\vskip 5truemm
In the following I review most of these recent works and discuss their
impact for cosmology. Though the review focuses on weak lensing, the distinction 
 between arclets and the weak lensing regime is somewhat arbitrary, and
 both are relevant for our purpose. Furthermore, since some of the results cannot be
discussed without referring to strong lensing, I often include paragraphs 
 which present new results from arcs and multiple image studies.  
  Section 2 quickly recalls the basic equations useful 
in gravitational lensing which help in the understanding of this 
 review.  The definitions for strong lensing cases  
 are not presented again, and I will refer to the 
 review by Fort \& Mellier (1994) for all these aspects.  
  In section 3, I focus on the mass 
distribution in clusters of galaxies from arc(let)s or mass
reconstruction from weak lensing inversion. I also address the issues
about the measurement of weak shear since it appears as a major challenge 
 for the observers. Section 4 presents  weak lensing
by large scale structures, and section 5 weak lensing by foreground galaxies on the 
 background sources (the so-called galaxy-galaxy lensing    
 analysis). I
then move toward the high-redshift universe in section 6. Section 7 and 8 
 are devoted to cosmological parameters and weak lensing on the
cosmological microwave background (CMB), respectively.  Conclusions and future 
prospects are discussed in the last section. 
\section{DEFINITIONS}
\subsection{Lensing equations}
\begin{figure}
\centerline{
\psdraft{figure=mellfig01.eps,width=14cm}}
\caption{\label{lensconfig.eps} {\it Description of a lensing configuration.} 
}
\end{figure}
\begin{figure}
\centerline{
\psdraft{figure=mellfig02.eps,width=13cm}}
\caption{\label{simulshearmap.eps} {\it Illustration of the two lensing regimes. The left panel is a simulation of a cluster of galaxies at redshift 0.15, 
modeled by an isothermal sphere with a velocity dispersion of 
1300 kms$^{-1}$.  The lensed population has an average redshift of one. 
 It the innermost region (bottom left part of the panel), tangential  and 
 radial arcs are clearly identified.
As the radial distance of lensed galaxies increases, the shear decreases, and far from the cluster
center, the ellipticity produced by the shear is lower than the intrinsinc ellipticity 
 of the galaxies. The lensing 
 signal must be averaged over a large number of galaxies in order to 
be measured
accurately. The zoom on the right panel shows the images of the galaxies in the 
 weak lensing regime. The contours show their shape 
as determined from their second moments. The average orientation of these 
galaxies is given by
the solid lines at the top right. The lower line is the true orientation
of the shear produced by the cluster at that position and the upper line
is the orientation computed from  92 galaxies of the zoomed area. The 
difference between the
two orientations is random noise due to the intrinsinc ellipticity and orientation 
 distributions of the galaxies.}  
}
\end{figure}
In this preliminary section, I do not discuss at length      
 the theoretical basis of the gravitational lens effect since
 all the details can be found in the comprehensive
textbook written by Schneider et al (1992).
 I focus on concepts and basic equations of the
gravitational lensing theory, in the thin lens approximation and for 
small deviation angles, which are necessary for this review.   \\
The apparent angular position of a lensed image,  $\thetag^I$ (in the
review, bold symbols denote vectors),
   can be expressed as function of 
  the (unlensed) angular position of the source,  $\thetag^S$, and the 
deflection angle, $\alphag(\thetag^I)$ as follows (see Figure 
\ref{lensconfig.eps}):
\begin{equation}
\thetag^I=\thetag^S+{D_{LS} \over D_{OS}} \alphag(\thetag^I) \ .
\end{equation}
$\alphag(\thetag^I)$ depends on the projected mass density of the lens, 
$\Sigma(\thetag^I)$,  and
the cosmological parameters through the angular-diameter distances from the lens $L$ to
the source $S$, $D_{LS}$,  from the observer $o$ to the source, $D_{OS}$
and from the observer to the lens , $D_{OL}$:
\begin{equation}
\alphag(\thetag^I)={4 \pi G \over c^2}D_{OL} \   \ {1 \over \pi} \int \Sigma(\thetag^I) {\thetag^I - \thetag' \over \vert \thetag^I - \thetag' \vert^2} d^2\theta' \ .
\end{equation}
$\Sigma(\thetag^I)$ can be expressed as a function of the 
Poisson equation and the strength of the lens is characterized by the ratio
 of the projected mass density of the lens to its critical projected
mass density $\Sigma_{crit}$ (see Fort \& Mellier 1994)  
\begin{equation}
{\Sigma(\thetag^I) \over \Sigma_{crit}}= {4 \pi G \over c^2}{D_{LS}D_{OL}\over
D_{OS}} \Sigma(\thetag^I)={1 \over 2}  \Delta \varphi(\thetag^I)
\end{equation}
where $\Delta$ is the 2-dimension Laplacian and $\varphi$ is the dimensionless gravitational potential projected
along the line of sight which is related to the projected gravitational
potential $\Phi$ as follows:
\begin{equation}
\varphi={2 \over c^2} \  {D_{LS}D_{OL}\over D_{OS}} \Phi \ .
\end{equation}
From differentiation of Eq.(1), we can express the deformation of an infinitesimal ray bundle 
as function of the Jacobian
\begin{equation}
{ d \thetag^S \over d \thetag^I}=A(\thetag^I) \ ,
\end{equation}
where $A(\thetag^I)$ is the {\it magnification matrix}:
\begin{equation}
A=\left(
\begin{array}{cc}
1-\partial_{11} \varphi & -\partial_{12} \varphi \\
-\partial_{12} \varphi & 1-\partial_{22} \varphi \\
\end{array}
\right)
\end{equation}
It can be written as function of two parameters (similar to the magnification and the
astigmatism terms in classical optics), the {\it convergence}, $\kappa$, and 
the {\it shear} components $\gamma_1$ and $\gamma_2$ of the complex
shear $\gammag=\gamma_1+i \gamma_2$:
\begin{equation}
A=\left(
\begin{array}{cc}
1-\kappa-\gamma_1 & -\gamma_2 \\
-\gamma_2 & 1-\kappa+\gamma_1 \\
\end{array}
\right)
\end{equation}
The isotropic component of the magnification, 
 $\kappa= 1/2 \ \Delta \varphi(\thetag^I)$, is directly related to the
projected mass density, and the two components $\gamma_1$ and $\gamma_2$
 describe an anisotropic deformation produced by the tidal gravitational
  field.  The eigenvalues of the magnification matrix are $1-\kappa \pm
\vert \gammag \vert $, where $\vert \gammag \vert =
\sqrt{\gamma_1^2+\gamma_2^2}$. They provide the elongation and the
orientation produced on the images of lensed sources.  The magnification of 
 an image is :
\begin{equation}
\mu = {1 \over det(A)} = {1 \over (1-\kappa)^2-\vert \gammag \vert^2} \ .
\end{equation}
The points of the image plane where $det(A)=0$ 
  are called the critical lines. The corresponding points of the source plane 
 are
called the caustic lines and produce infinite magnifiation (see Schneider et al
1992; Blandford \&
Narayan 1992; Fort \& Mellier 1994 for more detailed descriptions about
caustic and critical lines).
 The strong lensing cases correspond to configurations where 
sources are close to the caustic lines. These lenses have  
 $\Sigma(\thetag^I) / \Sigma_{crit} \geq 1$  and the convergence and shear are 
 strong enough to produce giant arcs and multiple images for suitably positionned 
 sources (Figures \ref{simulshearmap.eps} and  \ref{pnalarc.ps}). The weak 
lensing regime, which is the main topic of this review,  corresponds to lensing
configurations where $\kappa<<1$ and $\vert \gammag \vert <<1$.
In this regime, the magnification and the distortion of background
galaxies are so small that they cannot be detected on individual objects. 
   In that case, it is necessary to analyze  statistically the distortion 
 of the lensed population. \\
\subsection{Relation with observable quantities}
Let us assume that, to 
 first approximation, faint galaxies can be described as ellipses. Their
shape can be expressed as function of their weighted  
second moments which fully define the properties of an ellipse, 
\begin{equation}
M_{ij}= {\int S(\thetag) (\theta_i-\theta^C_i)(\theta_j-\theta^C_j) 
d^2\theta \over \int S(\thetag) d^2\theta } \ ,
\end{equation}
where the subscripts $i$  $j$ denote the axes (1,2) 
 of  coordinates $\theta$ in the 
 source and the image planes, $S(\thetag)$ is the surface brightness of the source and $\thetag^C$
is the center of the source. \\
Since the surface brightness of the source is conserved through the 
gravitational lensing effect (Etherington 1933),  
it is easy to show that, if one assumes that 
 the magnification matrix is constant across the image (lensed source), 
  the relation between
the shape of the source, $M^S$ and the lensed image, $M^I$ is
\begin{equation}
M^I=A^{-1} \ M^S \  A^{-1}
\end{equation}
Therefore, to first approximation, the gravitational lensing effect 
on a circular source changes its size (magnification) and transforms  
it into an ellipse (distortion) with axis ratio given by the ratio of the two
eigenvalues of the magnification matrix. The shape of the lensed galaxies 
can then provide information about these quantities. 
The approximation that the magnification matrix is constant over the
image area is always valid in the weak lensing regime,
because the spatial scale variation of the magnification is much larger  
than the typical size of the lensed galaxies (a
few arcseconds). This is not the case when the magnification tends to
infinity, but this case is beyond the scope of this review 
 (see Schneider et al 1992 and Fort \&
Mellier 1994). \\
The relation between the lens quantities described in the previous
section and the shape parameters of lensed galaxies is not immediate.  
Though $\gamma_1$ and $\gamma_2$ describe the anisotropic distortion 
 of the magnification, they are not directly related to
observables (except in the weak shear regime).  It is preferable to 
use the {\it reduced complex shear}, $\boldg$, and the complex 
polarization (or distortion), $\deltag$, which is an observable,
\begin{equation}
\boldg={\gammag \over (1-\kappa)} \ \ \ ; \ \ \ 
\deltag= { 2 g \over 1 + \vert \boldg\vert^2} ={2 \gammag (1-\kappa) \over
(1-\kappa)^2+\vert \gammag \vert^2} \ , 
\end{equation}
because $\deltag$ can be expressed in terms of the observed major and minor 
axes $a^I$ and $b^I$
of the image, $I$, produced by a circular source $S$:
\begin{equation}
{ a^2 - b^2 \over a^2 +b^2} = \vert \deltag \vert
\end{equation}
In this case, the 2 components of the complex polarization
are easily expressed with the second moments:
\begin{equation}
\delta_1={M_{11}-M_{22} \over Tr(M)} \ \ \ ; \ \ \ \delta_2={ 2 M_{12} \over Tr(M)} \ , 
\end{equation}
 where $Tr(M)$ is the trace of the magnification matrix. For non-circular sources, from Eq.(8) and Eq.(11) it is possible to relate the ellipticity of
the image $\epsilong^I$ to the ellipticity of the lensed source,
$\epsilong^S$. In the 
general case, it depends on the sign of $Det(A)$ (that is the position
 of the source with respect to the caustic lines) 
 which 
expresses whether images are radially or tangentially elongated. 
In most cases of
interest, $Det(A)>0$ (the external regions, where the weak lensing regime
applies) and :
\begin{equation}
\epsilong^I={1 + b^I / a^I \over 1 + b^I / a^I} e^{2 i \vartheta} =
{\epsilong^S + \boldg \over 1 - \boldg^* \epsilong^S}
\end{equation}
(Seitz \&  Schneider 1996), but when $Det(A)<0$:
\begin{equation}
\epsilong^I= {1+ \epsilong^{S*} \boldg \over \epsilong^{S*} + \boldg^*}
\end{equation}
These equations summarize most of the cases that will be discussed in this
review.  
\section{MASS DISTRIBUTION IN CLUSTERS OF GALAXIES}
\subsection{Mass reconstruction with arclets}
\subsubsection{Recent developments}
The use of arc(let)s and multiple images has already been discussed in
detail in Fort \& Mellier (1994). In the meantime, with the
refurbishment of the HST, spectacular images of arc(let)s and multiply imaged
galaxies have permitted enormous progress in this field.   It turns 
out that giant arcs are no longer the strongest constraints on 
cluster mass distribution, because 
 similar and even better information can be obtained with  
spatially resolved HST images of arclets. \\
The usual mass reconstruction technique using arc(let)s 
present in the innermost regions (close to the critical lines where
arcs are located) is based on the assumption that the cluster mass
density is smoothly distributed and can be expressed analytically, 
 possibly with addition of some substructures, and on the hypothesis  
that the observed arc(let)s correspond to rather generic lens
configurations, like fold, cups or lisp caustics.  These assumptions 
have already provided some convincing results using ground-based
images, in particular with predictions of the 
 position of additional images (counter images) associated with  
  arc(let)s  (Hammer \& Rigaud 1989, Mellier et
al 1993, Kneib et al 1993, 1995).  In all cases, it was found that
 the core radius of the dark matter distribution is small ($< 50
h_{100}^{-1}$ kpc) and that 
 its geometry is compatible with the faint extended
envelopes of light surrounding the giant cluster galaxies. 
 The new investigations using the detailed morphology of 
 the numerous  arc(let)s visible in the HST images (see Figure \ref{pnalarc.ps})  
have provided more refined constraints on the dark matter
 distribution on the 100 kpc-scale  
  (Hammer et al 1997, Gioia et al 1998,
 Kneib el al 1996, Tyson et al 1998).  
They confirm  
the trends inferred from previous ground-based data. \\
One critical issue about the approach described above is the possible 
 sensitivity of the result with the analytical mass profile used   
for the modeling. Since none of them has an unrealistic shape,  the global
property of the mass distribution on large scale is expected 
 to be rather well described. However, a direct comparison of the detailed mass
distribution with theoretical expectations seen in simulations is 
 difficult. Furthermore, the redshift distribution inferred from 
the lensing inversion (see section 5) can be strongly affected by 
the properties of the analytical model.  AbdelSalam, Saha \& Williams
(1998a,b) have recently proposed a non-parametric mass reconstruction 
algorithm which helps to overcome the limitation of analytical 
modeling. The technique uses arc(let)s with known redshifts as strong
constraints to recover a pixelised mass map of the lensing-cluster.
The pixel-mass reconstruction uses the smoothed projected light
distribution of the galaxy distribution 
  which is then pixelised exactly like the projected mass map. A 
  fit of a pixelised
Mass-to-Light ratio (M/L) permits to relate the projected light
distribution to the projected mass distribution for each pixel. The
results found for A370 and A2218 (for A2218, weak and strong lensing 
 features are used) are similar to those obtained otherwise, but this
approach looks a very interesting alternative which permits to have 
 a complete lens modeling based only on arc(let)s properties 
 (Dye \& Taylor (1998) attempted to generalize this approach 
 in order to compute the convergence and the shear, in the weak lensing regime).
\begin{figure}
\centerline{
\psdraft{figure=mellfig03.eps,width=13cm}}
\caption{\label{pnalarc.ps} {\it A panel of lensing  clusters observed
with HST. 
 The arc(let)s and multiple lensed images are indicated by a letter. In
A2390 (top left), the straight arc is made of two different galaxies corresponding
to images A and C.  The pairing of some images is obvious, like B in
A2390 (top-left), A in AC114, or A in A370.  Image B in MS2137 and B in A370 are 
radial arcs. A in MS2137 is a triple image from an almost
ideal configuration of a fold caustic.}
}
\end{figure}
\subsubsection{The X-ray/lensing mass discrepancy}
The use of X-ray
and optical images (ground-based or from the HST) of arc(let)s reveal  
 the  X-ray peaks are located at the center of the 
 most massive clumps of dark matter (Kneib et al 1995, Pierre et al 1996,  
AbdelSalam et al 1998a, Gioia et al 1998, Hammer et al 1997, Kneib el al 1996, 
 Nato et al 1998, Kneib et al in preparation). 
 On the other hand, the apparent contradiction
between the mass estimated from X-ray data and the lensing mass
($M_{lensing} \approx 2-3 M_X$),
 initially raised by Miralda-Escud\'e \& Babul (1995), is not totally   
clear. The puzzling results obtained on several clusters, sometimes on
 the same cluster but analyzed by different groups, have not yet provided 
conclusive statements about the mass density profile and the X-ray
versus dark matter dynamics.  B\"ohringer et al (1998) find an excellent
agreement between X-ray and lensing masses in A2390 which confirms 
 the  view claimed by Pierre et al (1996); Gioia et al (1998)
show that the disagreement reaches a factor of 2 at least in
MS0440+0204; Schindler et al (1997) find  a factor of 2-3  discrepancy 
for the massive cluster RX 1347.4-1145, but Sahu et al (1998) claim
that the disagreement is marginal and may not exist; Ota et al (1998) and 
 Wu \& Fang (1997) agree that there are 
 important discrepancies in A370, Cl0500-24 and Cl2244-02.\\
 There are still no definitive 
interpretations of these contradictory results. It could be that the 
 modeling of the gravitational mass from the X-ray distribution is not as simple. 
By comparing 
 the geometry of the X-ray isophotes of A2218 to the mass 
 isodensity contours of the reconstruction,   
Kneib et al (1995) found  significant discrepancies
in the innermost parts. The numerous substructures  visible in the 
 X-ray image have   
orientations which do not follow the projected mass density.
They interpret these features as shocks produced by the infalling X-ray gas, 
 which implies that the current  
 description of the dynamical stage of the inner X-ray gas is oversimplified 
(see  Markevich, 1997 and Girardi et al 1997 for similar views).  
Recent ASCA
observations of three lensing-clusters corroborate the view that
substructures are the major source of uncertainties (Ota et al  1998). \\
In order to study this possibility in more details,   
  Smail et al (1997) and Allen (1998) have performed a detailed comparison 
between the lensing mass and X-ray mass for a significant number of 
lensing clusters. Both works conclude that the substructures have a
significant impact on the estimate of 
X-ray mass. More remarkably, 
the X-ray clusters where cooling flows are present do not show a
significant discrepancy with X-ray mass, whereas the others X-ray
clusters do (Allen 1998).  This confirms that the discrepancy is certainly due to
 wrong assumptions on the physical state of the gas. These two  
 studies provide strong presumptions that we are now close to understand 
 the origin of the X-ray and lensing discrepancy.\\
An alternative has been
suggested by  Navarro, Frenk \& White (1997) who proposed that the 
analytical models 
 currently used for modeling mass distributions  may be inappropriate.  
Instead, they argue that the universal profile of the
mass distribution produced in numerical simulations of hierarchical
clustering may reconcile the lensing and X-ray masses. This is an
attractive possibility  because the universal profile is a natural outcome
from the simulations which does not use external prescriptions.
However, Bartelmann (1996) emphasized that the caustics produced by 
the universal profile predict that radial arcs should be 
thicker than observed in MS2137-23 
  (Fort et al 1993; Mellier et al 1994; 
Hammer et al 1997) and in A370 (Smail et al 1995), unless the sources 
 are very thin ($\approx 0.6$ arcsecond for MS2137-23).  
 This is not   a strong argument against the universal profile
  because this is possible
in view of the shapes of some faint galaxies observed with HST that some
diatnt galaxies are indeed very thin. But it
is surprising that no radial arcs produced by ``thick galaxies'' 
have  been detected so far. Even a selection bias would probably  
favor the observation of large sources rather than small thin and hardly
visible ones.  Evans \& Wilkinson (1998) have explored the range of 
 slopes of cups-like mass profiles which would permit to produce 
 radial arcs with thicknesses  as those observed.  As Bartelmann, they 
found that the universal profile does not work well, but that a more 
singular mass profile could be satisfactory. It is not mentioned,
however, whether these new profiles are compatible with the numerical
simulations of Navarro et al (1997).  \\
\subsubsection{Probing the clumpiness of clusters}
Besides,  HST images have also revealed the clumpiness of the
cluster mass distribution on small scales. Though most of the HST images 
of lensing-clusters shows arc(let)s with a coherent polarization on 
scales of 100 kpc, numerous perturbations are visible on scales   
 of about 10 kpc. The long-range pattern is
disrupted around most of the bright galaxies and show saddle-shape 
configurations as expected for clumpy mass distributions. In some extreme 
cases, giant arcs appear as broken filaments, probably
 disrupted by the halos around the brightest galaxies. With such amount of
details, one can therefore make a full mass reconstruction
 which takes into account all these clumps and possibly constrain 
 the mass of individual cluster galaxies. For giant arcs, this was
already stressed by Kassiola, Kovner \& Fort (1992), 
Mellier et al (1993), Kneib et al (1993), Dressler et al (1994),
 Wallington, Kochanek \& Koo (1995) and Kneib et al (1996). They used 
 the breaks (or the absence of breaks) in arcs 
 to put upper limits to the masses of a few cluster galaxies which are superimposed 
 on the arcs. The masses found for these cluster galaxies 
 range between $10^{10}$ M$_{\odot}$ and 2 $\times 10^{11}$ M$_{\odot}$, with
typical mass-to-light ratios between 5 and 15. \\
With the details visible on the 
HST images of arclets in A2218, AC114 or A2390, 
 the sample of halos which can be constrained by this 
 method is much larger and can provide more 
significant results. The number of details permits also to 
 use more sophisticated methods of investigation.   
 The most recent procedure uses the galaxy-galaxy lensing 
analysis. This technique is described in Section 5, but since the 
clumpiness of dark matter in clusters is strongly related to the 
halos of cluster galaxies,  
 I present the use 
of  galaxy-galaxy lensing for cluster galaxies in this section. \\
 The simplest strategy is to 
start with an analytical potential which reproduces the general
features of the shear pattern of HST images, and in a second step, to include 
in the model  analytical 
halos around the brightest cluster members. In practice, additional mass components 
 are put in the model in order to interpret the arc(let)s which cannot be
easily explained by the simple mass distribution.  Some guesses are done 
in order to pair unexplained  multiple images. The colors of 
the arc(let)s as well as their 
morphology help a lot to make these associations.  
This approach has been  proposed   
 by Natarajan \& Kneib (1997), and Natarajan et al (1998). The detailed study done
in AC114 by Natarajan et al 
 indicates that about 10\% of the dark matter is associated with 
 halos of cluster galaxies. These halos 
have truncation radii smaller than field galaxies ($r_t \approx $15kpc) 
with a general trend 
of S0-galaxies to be even more truncated than the other galaxies.  If this result is confirmed it would
be a direct evidence that truncation by tidal stripping is really 
efficient in rich clusters of galaxies.  This result is somewhat
contradictory with the absence a clear decrease of rotation curves 
 of spiral galaxies in nearby clusters (Amram et al 1993) which 
 is interpreted as a proof that massive halos of galaxies still exist 
 in cluster galaxies. However,
it could be explained if the spirals which have 
been analyzed appear to be in the cluster center only by projection effects
 but  are not really located 
in the very dense region of the clusters where stripping is 
 efficient. \\
Geiger \& Schneider (1997, 1998) used a maximum likelihood
analysis which explores simultaneously the distortions induced by the
cluster as a whole and by its individual galaxies.  They applied this analysis 
 to the HST data of Cl0939+4713 and reached  similar conclusions 
 as Natarajan et al (1998).  Several issues limit the reliability of their 
analyses and of the other methods as well (Geiger \& Schneider 1998).  
First, depending on the slope 
of the mass profile of the cluster, the contributions of the cluster 
 mass density and of the cluster galaxies  may be difficult to 
 separate.  Second, it is necessary to have a realistic model for 
 the redshift distribution of the background and foreground 
galaxies.  Finally, the mass sheet degeneracy (see Section 3.2) is also an additional 
source of uncertainties.  Regarding these limitations, Geiger \&
Schneider discuss the capability of the 
galaxy-galaxy lensing in clusters to provide valuable constrains 
 on the galactic halos from  the data they have in hands.  Indeed,
  some of the issues they 
raised can be solved, like for instance the redshift distribution 
of the galaxies. It would be interesting to look into more details 
 how the analysis could be improved with more and better data. 
\begin{figure}
\centerline{
\psdraft{figure=mellfig04.eps,width=16cm}}
\caption{\label{comparealgo.ps} {\it Examples of different algorithms for
the mass reconstruction of clusters. Image A (top left) is the original
simulated lensing cluster. Panel B shows an original Kaiser \& Squires
(1993) mass reconstruction, assuming that all background sources are circular.
 Panel C shows the same reconstruction as panel B, but with another
smoothing method, which was proposed by Seitz \& Schneider (1995), in 
 order to smooth the distortion distribution. The panel D shows the same
result but the sources have an ellipticity distribution. Panel
E shows the same reconstruction as panel D, but uses an adaptive
smoothing scale. In panel F the linear and non-linear weak lensing
regimes are now used in the inversion. Panel G shows the same
reconstruction algorithm as F, with an  additional extrapolation 
 of the distortion field outside the field. The last panel shows 
 the mass reconstruction with the constraint that the minimal
 mass density at any point is zero (from Seitz \& Schneider 1995a). 
 } 
}
\end{figure}
\subsection{Mass reconstruction from weak lensing}
A powerful and complementary way to recover the mass distribution 
of lenses has been proposed by Kaiser \& Squires (1993). It is based on
the distribution of weakly lensed galaxies rather
than the use of giant arcs. In 1988, Fort et 
al (1988) have obtained at CFHT deep sub-arcsecond images 
 of the lensing-cluster A370 and 
 observed the first weakly distorted galaxies ever detected. 
 The galaxy number density of their observation was about 
 30 arcmin$^{-2}$, mostly composed of background sources, far beyond the 
 cluster. These galaxies lensed by the cluster show a correlated 
 distribution of ellipticity/orientation which maps the projected mass 
density. The  first attempt to use this distribution 
of arclets as a probe of dark matter has been done by Tyson et al   
(1990), but the theoretical ground and a rigorous inversion technique
was first proposed by Kaiser \& Squires.   \\
By combining Eq.(2),(4) and (5), one can express the complex shear as 
a function of the convergence, $\kappa$ (see Seitz \& Schneider 1996 and references
therein):

\begin{equation} 
\gammag(\thetag) = {1 \over \pi} \int 
\mathcal{D}(\thetag-\thetag') \ \kappa(\thetag') \ d^2\theta'  \ ,
\end{equation}
where 
\begin{equation}
\mathcal{D}(\thetag-\thetag')={(\theta_2-\theta_2')^2 -
(\theta_1-\theta_1')^2- 2i
(\theta_1-\theta_1')(\theta_2-\theta_2') \over
\vert (\thetag-\thetag') \vert^4 } \ .
\end{equation}
This equation can be inverted in order to express the
projected mass density, or equivalently $\kappa$, as function of the shear:
\begin{equation}
\kappa(\thetag)= {1 \over \pi}  \int \Re[\mathcal{D}^*(\thetag-\thetag')
\gammag(\thetag')]  \ d^2\theta' \ + \kappa_0  \ ,
\end{equation}
where $\Re$ denotes the real part.   
From Eq.(14) we can express the shear as a function of the complex
ellipticity. Hence, if the background ellipticity distribution is randomly
distributed, then $<\vert \epsilong^S \vert>=0$ and 
  
\begin{equation}
<\vert \epsilong^I \vert>=  \vert \boldg \vert = {\vert \gammag \vert
 \over 1 - \kappa} 
\end{equation}
 (Schramm \& Kayser 1995). In the most extreme case, when $\kappa<<1$ (the linear regime 
 discussed initially by Kaiser \& Squires 1993), 
 $<\vert \epsilong^I \vert> \approx  \vert \gammag \vert$, and 
 therefore, the projected mass density can be recovered directly from the
measurement of the ellipticities of the lensed galaxies.  \\ 
The first cluster mass reconstructions using the  
  Kaiser \& Squires linear inversion have been done by Smail (1993) and 
Fahlman et al (1994). Fahlman et al estimated the total mass within a circular radius
using the {\it Aperture densitometry} technique (or the ``$\zeta$-{\it statistics}''), which 
consists in computing the difference between 
 the mean projected mass densities within a radius $r_1$ and 
  within an annulus $(r_2-r_1)$ (Fahlman et al 1994, Kaiser 1995) as
function of the {\it tangential shear}, $\gamma_t=\gamma_1 {\rm
cos}(2\vartheta)+\gamma_2 {\rm sin}(2\vartheta)$ (see Eq.(14)), averaged in the ring:
\begin{equation}
\zeta(r_1,r_2)=<\kappa(r_1)>-<\kappa(r_1,r_2)>={2 \over 1-r_1^2/r_2^2} \ \int_{r_1}^{r_2} <\gamma_t>
 d {\rm ln}r \ .
\end{equation}
This  quite robust mass estimator minimizes
 the contamination by foreground and cluster galaxies and permits a simple check that the
signal is produced by shear,  simply by changing $\gamma_1$ in $\gamma_2$ and $\gamma_2$
in $-\gamma_1$ which should cancel out the true shear signal.\\
The mass maps inferred from their images coincide with  
 the light distribution from the galaxies. But,   the impressive
 M/L found in the lensing cluster MS1224+20 by Fahlman et al (see Table 1)
  led to a surprisingly high
value of $\Omega$ (close to 2!).  This result is somewhat 
 questionable and is probably due to the various
sources of errors, possibly in the correction of PSF anisotropy (see
Section 3.3). Furthermore, 
 the  2-dimension mass reconstructions presented in the very first
papers  looked noisy, probably 
 because of  boundary effects due to the
intrinsically non-local reconstruction, the geometry of the finite-size
 charge coupled divice (CCD), and the reconstruction algorithm 
can have terrible effects on the
inversion. These problems have been discussed in several papers 
(Schneider \& Seitz 1995, Seitz \&
Schneider 1995a, Schneider 1995, Seitz \& Schneider 1996). In particular, 
Kaiser (1995) and
Seitz \& Schneider (1996) generalized the inversion to the non-linear
regime, by solving the integral equation obtained from Eq.(18) by replacing 
 $\gammag$ by $(1-\kappa)\boldg$, or 
similarly  by 
 using the fact that both $\kappa$ and $\gammag$
 depend on second derivatives of the projected gravitational potential 
$\varphi$ (Kaiser 1995) which permits to recover the mass density by 
 this alternative relation:
\begin{equation}
\nabla {\rm log}(1-\kappa)= {1 \over 1 -\vert \boldg \vert^2} 
 \ . 
\left(
\begin{array}{cc}
1+g_1 & -g_2 \\
-g_2   & 1-g_1 \\
\end{array}\right) \  \
\left(
\begin{array}{c}
\partial_1 g_1 + \partial_2 g_2 \\
\partial_1 g_2 - \partial_2 g_1 \\
\end{array}
\right)
\end{equation}
Both Eq.(18) and Eq.(21) express the same relation between $\kappa$ and 
 $\gammag$ and can be used to reconstruct the projected mass
density.  \\
The improvements which   
have been proposed and discussed in detail by Seitz \& Schneider (1995a), 
Kaiser
(1995), Schneider (1995), Bartelmann (1995c), Squires \& Kaiser (1996), 
Seitz \& Schneider (1996, 1997) or Lombardi \& Bertin (1998a,b) lead to reliable   
mass reconstructions from lensing inversion, and comparison with simulated 
  clusters  proves that it can be     
  considered now as a robust technique (see Figure \ref{comparealgo.ps}, in particular
the comparision between panel A and panels G and H).  However, the
mass distribution recovered is not unique because the addition of a lens
plane with constant mass density will not change the distortion of the
galaxies (see Eq.(18)). Furthermore, the inversion only uses the ellipticity of the
galaxies regardless of their dimension, so that changing  $(1-\kappa)$ in
$\lambda(1-\kappa)$ and $\gammag$ in $\lambda$ $\gammag$ keeps $\boldg$ 
invariant.
This so-called {\it mass sheet degeneracy}  
 initially reported by Gorenstein, Falco \& Shapiro (1988), has been 
 pointed out by 
 Schneider \& Seitz (1995) as a fundamental limitation of the lensing 
 inversion. \\
The degeneracy could in principle be broken
 if the magnification can be measured independently, since 
 it is not invariant under the linear transformation mentioned above,
but instead it is reduced but a factor $1/\lambda^2$. 
Broadhurst, Taylor \& Peacock (1995) have proposed to measure directly 
 the magnification by  using the 
 magnification bias which changes the galaxy number-counts (see Section
3.4),  whereas Bartelmann \&
Narayan (1995) explored their {\it lens parallax method} which compares 
 the angular sizes of lensed galaxies with an unlensed sample.  The
 lens-parallax method requires a sample of unlensed population 
 having the same surface brightness distribution. However,  
  in the case of  ground-based observations, for the 
faintest (i.e. smallest galaxies) the convolution of the signal by 
the seeing disk can significantly affect the measurement of their surface 
brightness. 
Therefore, the method needs a careful 
 handling of small-size objects.  A more promising appraoch is the use of
wide field cameras with a typical field of view much larger than clusters 
of galaxies. In that case  $\kappa$ should  vanish at the boudaries of the field, so that the 
degeneracy could in principle be broken.\\
An attractive alternative has been suggested by Bartelmann et al (1996)
who proposed a maximum likelihood reconstruction algorithm. The 
advantage  is that this is a {\it local} approach since 
 it fits the projected potential at each grid-point.  The observables
 are constrained by the second-derivatives of the potential, 
 using a least-$\chi^2$ which   
 computes simultaneously both the magnification and the distortion which 
 are compared to the ellipticity and the sizes of the galaxies.  
 Squires \& Kaiser (1996) and Bridle et al (1998) have investigated
similar maximum-likelihood techniques with different regularizations, though  
they fit the projected mass density rather than the potential. 
 It seems however more attractive to use 
the deflection potential rather than the projected mass distribution
in order to avoid the incomplete knowledge of the contribution to the
projected mass density of the matter outside the observed area 
 (Seitz et al 1998).\\
\begin{figure}
\centerline{
\psdraft{figure=mellfig05.eps,width=14.5cm}}
\caption{\label{massa2218.eps} {\it Weak lensing analysis and mass
reconstruction of A2218 (from Squires et al 1996a). The images have been
 obtained at CFHT in I-band. The top-left panel shows the smoothed galaxy number
density and the top-right shows the smoothed light distribution. The bottom-left is the
shear map. The length of each line is proportional to the amplitude of the shear. From
this shear map, the mass reconstruction of the Kaiser \& Squires (1993) algorithm
produces the mass map of the bottom-right. The correlation betwenn the light and the
matter distribution is clear.} 
}
\end{figure}
The comparison  done by Squires \& Kaiser (1996) between the {\it direct
reconstructions}, like the Kaiser \& Squires (1993) approach, and the 
{\it inversion methods}, as those maximum likelihood reconstructions
 alternatives did not lead to conclusive results, though the maximum
likelihood inversion looks somewhat better.  
  But it is worth pointing out  that one of the 
advantages of the maximum likelihood inversion is that 
  it eases the addition of some observational constraints, 
 like strong
lensing features (Bartelmann et al 1996; Seitz et al 1998).  
 More recently, Seitz et al (1998) 
 proposed an improved entropy-regularized maximum-likelihood inversion
where they no longer smooth the data, but they use 
 the ellipticity of each individual galaxy. \\
\begin{table}
{\small
\caption{Main results obtained from weak lensing analyses of 
lensing-clusters. The 
scale is the typical radial distance with respect to the cluster center.
The last cluster has two values for the M/L ratio. This corresponds to
two extreme redshifts assumed for the lensed population, either $z=3$ or
$z=1.5$. For this case, the two values given for the velocity dispersion
are those inferred when $z=3$ or $z=1.5$ are used. }
\bigskip
\begin{tabular}{lccccccl}\hline\hline \\
\bigskip
Cluster& $z$ & $\sigma_{obs}$ &  $\sigma_{wl}$ & M/L & Scale& Tel. & Ref. \\
 & & (kms$^{-1}$)  & (kms$^{-1}$)& ($h_{100}$) & ($h^{-1}_{100}$ Mpc) & &  \\
\hline
A2218 & 0.17 & 1370 &-& 310 & 0.1 & HST & Smail et al (1997)\\
A1689 & 0.18 &2400&  1200-1500 & - & 0.5 & CTIO & Tyson et al (1990)\\
      &  & &-& 400& 1.0 & CTIO & Tyson \& Fischer (1995)\\
A2163 & 0.20 & 1680&740-1000  & 300&0.5 &  CFHT & Squires et al (1997)\\
A2390 & 0.23 & 1090& $\approx$1000 &320 & 0.5&CFHT & Squires et al (1996b) \\
Cl1455+22 & 0.26  &$\approx$ 700 & - &1080&0.4 & WHT & Smail et al (1995)\\
AC118 & 0.31 & 1950 & -&370 & 0.15 &HST  &  Smail et al (1997)\\
Cl1358+62 & 0.33  & 910&780 &180 & 0.75&HST  & Hoekstra et al (1998)\\
MS1224+20 & 0.33  & 770& -& $\approx$ 800& 1.0& CFHT & Fahlman et al (1994) \\
Q0957+56  & 0.36  & 715&- & -& 0.5& CFHT& Fischer et al (1997) \\
Cl0024+17 & 0.39  & 1250 &-& 150& 0.15 & HST &  Smail et al (1997)\\
          & & & 1300& $\approx$900& 1.5& CFHT & Bonnet et al (1994)\\
Cl0939+47 & 0.41  & 1080 &-&120 & 0.2 & HST &  Smail et al (1997)\\
          &    &  &-&$\approx$250 & 0.2& HST& Seitz et al (1996) \\
Cl0302+17 & 0.42  & 1080& & 80& 0.2 & HST &  Smail et al (1997)\\
RXJ1347-11 & 0.45  & - &1500& 400& 1.0 & CTIO& Fischer \& Tyson (1997)\\
3C295 & 0.46 & 1670 &1100-1500 &- &0.5 & CFHT & Tyson et al (1990)\\
 &  & &-& 330&0.2 & HST &   Smail et al (1997)\\
Cl0412-65 & 0.51  &-&-&70 & 0.2& HST & Smail et al (1997) \\
Cl1601+43 & 0.54  &1170 &-&190 & 0.2& HST & Smail et al (1997) \\
Cl0016+16 & 0.55  &1700 &-&180 & 0.2& HST &  Smail et al (1997)\\
          &       & &740 &740&0.6 & WHT & Smail et al (1993)\\
Cl0054-27 & 0.56  &-&-&400 & 0.2& HST &  Smail et al (1997)\\
MS1137+60 & 0.78  &859$^1$&-&270 &0.5& Keck&  Clowe et al (1998)\\
RXJ1716+67 & 0.81  &1522$^2$&-&190 &0.5 &Keck & Clowe et al (1998)\\
MS1054-03 & 0.83 & 1360$^3$&1100-2200 &350-1600 &0.5 & UH2.2& Luppino \& Kaiser (1997)\\
\\

\hline
\end{tabular}
}
{\small $^1$ Gioia, private communication. \ \ $^2$ Gioia et al (1998)
 \ \ $^3$ Donahue et al (1998).}
\end{table}

Since 1990, many clusters have been investigated using the weak lensing
inversion, either using ground-based or HST data. They are summarized in Table (1), 
 but the comparison of these results is not straightforward because of
the different observing conditions which produced  each set of data 
 and the different mass reconstruction algorithms used by each author.  
 Nevertheless, all these studies
show that on scales of about 1 Mpc, the geometry of 
mass distributions, the X-ray
distribution and the galaxy distribution are similar (see Figure \ref{massa2218.eps}), 
though the ratio
of each component with respect to the others may vary with radius. The
 inferred M/L ratio lies between 100 to 1600, with a median
value of about 300, with a trend to increase with radius. 
   Contrary
to the strong lensing cases, there is no evidence of discrepancies
between the X-ray mass and the weak lensing mass. It is worth 
 noting that the strong lensing mass and the weak lensing mass estimates
 are consistent in the region where the amplitude of two regimes 
 are very close. This is an indication 
that the description of the X-ray gas, and its coupling 
 with the dark matter on the  
scales corresponding to strong lensing studies is oversimplified,
whereas on larger scales, described by weak lensing analysis, the
detailed description of the gas has no strong impact. \\
The large range of M/L could partly be a result of one of the 
issues of the mass reconstruction from weak lensing.  As shown in
Eq.(1), the deviation angle depends on the ratios of the three 
angular-diameter distances, which varies with the redshift
assumed for the sources. For low-z lenses, the dependence with redshift
of the background galaxies   
is not considerable, so the calibration of the mass can be provided with
a reasonable confidence level.  However, distant clusters
 are highly sensitive to the redshift of the sources, 
  and it becomes very difficult to scale the total
mass without this information, even though the shape of the projected 
mass density is
reconstructed correctly.  The case of high-redshift 
clusters is really more complicated. For a low-$z$ cluster (say $z<0.4$), 
it is not necessary to go
extremely deep since the background galaxies are between $z$=0.4 to 1. 
 So spectroscopic surveys can provide the redshift distribution with a
 good accuracy.  In contrast,  the background sources lensed by 
 high-$z$ clusters 
 are beyond $z=1$ are therefore are dominated by very faint 
 galaxies  ($I>22.5$) which cannot be observed easily by spectroscopy.
\vskip 5truemm
The masses inferred from the strong lensing and weak lensing
reconstruction   put valuable constraints on the median 
 M/L of lensing clusters. From the investigation of
about 20 clusters, the median M/L is lower than 400.  
This implies that weak  lensing analyses predict  
   $\Omega<0.3$ with a high significance level. Even if 
the uncertainties are large and if the weak lensing inversion needs 
to be improved, the HST data, in particular for clusters with giant arcs and 
many arc(let)s with known redshift, imply that the mass of clusters of 
galaxies cannot be reconciled with an Einstein-de Sitter (EdS) universe. The constraints 
 on $\Omega$ are in good agreement with other observations (see 
 the recent discussion by Krauss 1998).  \\
Another strong statement results from the mass reconstruction obtained 
by Luppino \& Kaiser (1997) and Clowe et al (1998) or the detection 
 of giant arcs in very distant clusters (Deltorn et al 1997): massive clusters 
 do exist at redshift $\approx$1 ! Though the total mass and the M/L cannot be 
 given with a high accuracy, it cannot change the conclusion, unless 
 unknown important systematics have been neglected. 
Therefore, we now have the first direct observational  evidences 
 that high mass-density peaks have generated massive clusters of
galaxies at redshift 1. These results are extremely promising 
 and  are corroborated by other weak lensing studies around radiosources
and quasars (see section 4.2). Indeed, since they already  question the 
standard CDM model and rather favor low-density universes,  we   
can certainly expect fantastic developments of the investigation of 
high-redshift clusters with weak lensing during the coming years.
\\
The impressive and spectacular results obtained from weak lensing 
have demonstrated the power of this technique. However, though the
results seem reliable in the cluster center (say within 500 kpc), 
there are still uncertainties outward, when the shear becomes very 
small. 
A critical issue of the lensing inversion is the 
reliability of the mass reconstruction and how it degrades when the shear
decreases.
Kaiser (1995) emphasized that Eq.(20) can be used as a check of the 
mass reconstruction since the curl of $\nabla {\rm log}(1-\kappa)$
should be zero only if the  shear is  recovered properly.   Van Waerbeke (1999) 
 has recently proposed an elegant way to estimate the accuracy of the mass 
reconstruction from the noise properties of the reconstruction.   
 Nevertheless, the comparison of results using different algorithms and the 
 stability and the reproducibility of each inversion have still 
 to be done in order to  demonstrate  
that weak lensing analysis produces reliable results. This is 
 impportant  for
the future, when ground based observations of very large fields will be
 performed.  Mellier et al (1997) have compared the shear maps obtained
in A1942 by using the Bonnet-Mellier  and the ACF methods (see 
 next section). Though the two maps
are similar, discrepancies are  visible at the periphery, 
  but no quantitative estimates of the similarity of both 
 maps are given.  An
important step has been done by Van Waerbeke et al (in preparation)  who 
 analyzed A1942  
 using different shear measurements, different mass reconstruction 
algorithms and different data, obtained with two CCD cameras mounted at
CFHT. Three sets of data have been used, all of them having a total 
exposure time of 4 hours and a seeing of 0.7". The results show 
impressive similarities even in the details of the mass reconstruction,
down to a shear amplitude of 2\%.  This is the first work which demonstrates 
that results are stable and are not produced by artifacts, even at a 
 very low shear amplitude.  
 The main concerns are the capabilities of instruments, and 
the image analysis algorithms to measure very weak shear. This critical
issue deserves a detailled discussion, which is the subject of the next
section.
\subsection{Measuring weak shear}
In addition to the technical problem of the mass reconstruction
algorithm and the redshift distribution of the sources, weak lensing
 is also sensitive to the accuracy of the measurement of 
 ellipticities of lensed galaxies. The atmosphere  has  
 dramatic effects, in particular the seeing 
circularizes the innermost part of galaxies which 
 affects the  measurements of shapes of faint
galaxies (see for example the simulations by Bartelmann 1995c). 
These issues have been investigated in details by 
 Bonnet \& Mellier (1995), Mould et al (1995), Kaiser, Squires \& Broadhurst 
 (1995) and 
Van Waerbeke et al (1997).   Atmospheric
 effects (seeing, atmospheric refraction, atmospheric dispersion), 
 telescope handling (flexures of the telescopes, bad guiding) and optical 
distortions are extrinsic problems which can bias the measurements, 
 though in principle they can be corrected in various
ways. The atmospheric dispersion can be minimized by using a I-band
filter and by observing clusters close to zenith, which  also
minimizes the flexures. Optical distortions can be corrected either 
analytically, if the optics is known perfectly or by using the 
stars located in the fields. On the other hand, the ellipticity distribution 
of the galaxies is an intrinsic source of noise. \\
The extrinsic and
intrinsic noises compete together: the circularisation 
by the seeing is important only for the faint galaxies because their
typical size is of the same order as the seeing disk. So, in principle
it should be better to use big (bright) galaxies only, though they are not 
 as numerous as the faint (small) galaxies.  On the other hand, 
 the noise produced by the intrinsic ellipticity distribution of the
galaxies is minimized by averaging the shape of a large number of 
galaxies.  The typical scale on which galaxies can be averaged is 
 defined by the spatial resolution of the reconstructed mass map. For 
 intermediate and high redshift clusters of galaxies
 the typical angular scale is a few arcminutes, so that galaxies must be 
averaged on
less than one arcminute in order to map the projected mass-density with a good sampling.  
If we assume an ellipticity
dispersion, $\sigma_{\varepsilon}$ of about 0.3, as it is suggested by 
nearby surveys and the 
 distribution of galaxies in the Hubble Deep Field (HDF), then we can measure
an ellipticity, $\epsilon$, produced by a gravitational shear of $\vert \gammag \vert$=10\% if 
the number of averaged galaxies, $N$ is:
\begin{equation}
N > \left({\sigma_{\varepsilon} \over \epsilon}\right)^2 \approx 10 \ .
\end{equation}
It it therefore quite easy to measure gravitational shear of 10\% on 
arcminute scale. But going down to 1\%  would require about 
 900 galaxies which is not feasible  on such 
angular scale, unless many fields of one arcminute are averaged. An 
alternative is to go deeper 
 in order to increase the galaxy number density. 
 But this is not sufficient to 
 increase the accuracy of the results, because most of the faint galaxies
 have unknown redshift distribution which prevents to scale the 
mass properly, and also because it is difficult to correct them from PSF. \\
The procedure  to recover the shear field    
 from the ellipticities of individual galaxies has several 
solutions.  Bonnet \& Mellier (1995) compute the second moment of galaxies 
 (see Eq.(9)) within a circular annulus and average the signal on a given area  
 (a {\it superpixel}) by only using the faint galaxies which dominate
their deep observations. The size of the 
 inner radius is constant and close to the seeing
disk, which minimizes the effect of the circularisation of the innermost 
isophotes on the measurement of the ellipse.  The outer radius is 
 also constant and has been
optimized by using simulations of galaxies in order to get the highest   
signal-to-noise ratio on each second moment. The drawback of this approach is
that the second moment of this annulus is no longer a direct measurement of 
the shape 
 of the galaxy and it must be calibrated by simulations
 for each observing conditions. The anisotropy of the point spread function (PSF) 
is corrected  
on the averaged signal, 
 assuming it is dominated by optical defects and that it  behaves
like a stretching of the image. This assumption is not valid for  
individual galaxies, but for each superpixel, deep observations average
 so many galaxies that it is possible to assume that the resulting 
 signal reflects the one produced by an ideal galaxy having the same
profile on each superpixel. In this case,  Bonnet \& Mellier have shown
from simulations that the correction works very well down to a shear
amplitude of 3\% (see Bonnet et al 1995, and Schneider et al 1998a).\\
Kaiser et al (1996) (see also Luppino \& Kaiser 1997, Hoekstra et al 1998 for 
 further developments)
 compute the second moment within a variable aperture which 
depends on the size of individual galaxies; but instead of an annulus 
 they use a Gaussian filter, and introduce  
 a more rigorous correction of the PSF anisotropy. Since they 
 do not make selection from the size of the galaxies  to measure the
shear, it is clear that the biggest galaxies require less correction.  
 Therefore their correction depends on the total area of each individual galaxy.
Assuming the anisotropy of the PSF is small, 
 Kaiser et al introduced  a smearing correction,  
defined by a  linear smear polarizability which expresses the (small) shift of 
polarization of galaxies induced by an anisotropic PSF. To first order, this shift 
can be expressed
analytically and provides the correction  from the shapes of the stars visible 
in the field, by dividing the smear polarizability by the observed polarization of
stars.  The  efficiency of this method has been tested by using      
  HST images which were degraded to the corresponding
 PSF anisotropy expected on ground-based images.  They proved that
  the correction works very well. 
  However, the method only works for bright
galaxies. For fainter samples, they calibrated the polarization-shear relation by
artificially lensing HST images and then by degrading them by the PSF observed in their
data.  However, it seems more preferable to calibrate the PSF anisotropy 
 directly from the images. 
 Luppino \& Kaiser (1997) calibrated the anisotropic correction only from the 
 observations of the stars in their fields, without auxiliary data. \\
 Mould et al (1995) proposed
a different procedure: they compute the second moment within a limiting
isophote rather than a finite aperture and corrected linearly from the 
PSF anisotropy assuming, as Kaiser et al, that the correction 
 is inversely proportional to the  area of the source.\\ 
Van Waerbeke et al (1997) have proposed an original alternative which 
fully exploits the signal down to the noise level on each CCD images and
 reduces the error of the second moment of galaxies. Instead of using
individual objects, they compute the local auto-correlation 
function (ACF), $\xi(\thetag)$, of pixels, averaged on a given area.
 The (unlensed)  ACF of the averaged source population, $\xi^S$, is an 
isotropic quantity, which by definition 
 is centered, and therefore does not depend on the detection 
procedure and on the computation algorithm of its centroid.  
 When a  lensing signal 
is present, the shape of the ACF of the 
 lensed population, $\xi^I(\thetag)=\xi^S(A\thetag)$ is no longer isotropic. For
example,   in the weak lensing regime it writes:
\begin{equation}
\xi^{I}(\thetag)=\xi^{S}(\vert \thetag \vert)- \vert \thetag \vert \ \partial_{\thetag}
 \xi^{S}(\vert \thetag \vert) \ \left(I-A \right)  \ ,
\end{equation}
where $I$ is the identity matrix. The ACF is now composed of the unlensed isotropic 
component plus an anisotropic 
term which depends on the magnification matrix, $A$,  which  stretches it   
like a real object. The second moment of the ACF    
 can also be expressed as a function of the distortion and the
magnification (Van Waerbeke et al 1997). Since all the pixels of the image are 
used, the ACF uses the full
information of the image; in particular the flux coming from extremely
faint objects, for which the measurement of a centroid and the second
moments are not measurable precisely, is also taken into account.  For 
that reason, in principle, the ACF can work on
images which reach the confusion limit (Van Waerbeke et al 1997, R\'efr\'egier \&
Brown 1998).  It turns out that  in practice it is better to
 use the ACF around detected galaxies rather on
the total image because correlated noises like 
 electronics cross-talk or shift-and-add residuals, may generate 
 spurious coherent signal (Van Waerbeke \& Mellier 1997). This method looks ideal for the 
optimal 
 extraction of weak lensing signals because the signal-to-noise ratio of
the ACF is always high enough and  spreads over sufficient pixels 
 to avoid the need of circular filtering, like faint galaxies, and 
 to provide an accurate estimate of its shape parameters .\\
The reliability of the relation between measured ellipticities and shear,
and of the mass reconstruction obtained from  observations has been checked by
 Kaiser \& Squires (1993), Bonnet \& Mellier (1995), and by independent
simulations of Bartelmann (1995c) and Wilson, Cole \& Frenk (1996a). 
Despite careful studies to check whether images are
corrected accurately from circularisation by seeing and 
PSF anisotropy (in particular the spatial variation of the PSF in the 
field can be modeled), there is still a lot of work to do in this area. 
 For instance, the
weighting functions proposed to measure the ellipticities is based on intuition
but no complete investigation has been done 
so far in order to find the optimal 
 one.  Moreover, for each of these procedures, it is assumed that 
the PSF anisotropy in unidirectional. This may be not true, in particular
when instruments with poor optical design are used.  In that case, 
 the correction becomes non trivial, and paradoxically this could appear 
on the image with the best image quality because details of the PSF are 
no longer smeared by circularisation of the seeing.  This domain is 
certainly at its
infancy and much work and new ideas should appear during the next
years, mainly because the shear produced by large scale structures
 is expected to be very small.
\subsection{Mass profile from the magnification bias}
In parallel to the mass reconstruction using weak shear measurements, 
 one can   use the direct measurement of the magnification 
 from  the local modification of the galaxy 
number density. This ``magnification bias'' expresses 
the simultaneous effects of  the gravitational magnification, which 
 increases the flux received from any lensed galaxies and
permits the detection of galaxies enhanced by the amplification, 
but also magnifies by the same amount the area of the projected 
lensed sky and thus decreases the apparent galaxy number density.
The total amplitude of the magnification bias
depends  on the slope of
the galaxy counts as  a function of magnitude and
on the magnification factor of the lens. For a circular lens, the 
 radial galaxy number density of background galaxies writes:
\begin{equation}
   N(<m,r) = N_0(<m) \ \mu(r)^{2.5\alpha-1} \ \approx N_0 \ (1+2
\kappa)^{2.5\alpha-1} \ \ \ \  {\rm if } \ \kappa\ {\rm and} \ \vert
\gammag\vert \ll 1\  ,
\end{equation}
 where  $\mu(r)$ is the magnification, $N_0(<m)$ the
intrinsic (unlensed) number density, obtained from 
 galaxy counts in a nearby empty field, and
  $\alpha$  is the intrinsic count slope:
\begin{equation}
\alpha = {{\rm d}logN(<m) \over {\rm d}m }\ .
\end{equation}
A radial magnification bias $N(<m,r)$ shows up only when
the slope $\alpha \not=0.4$; otherwise, the increasing number of
magnified sources is exactly compensated by the apparent field dilatation.
For slopes larger than $0.4$ the magnification bias increases the galaxy
number density, whereas for slopes smaller than $0.4$ the radial density
 will show a depletion.
Hence, no change in the galaxy number density can
be observed for $B(<26)$ galaxies, since
the slope is almost this critical  value (Tyson 1988).
 But it can be detected in the $B>26$, $R>24$ or $I>24$ bands when the slopes 
are close to 0.3 (Smail et al 1995).  \\
The change of the galaxy number density can be used as a direct
measurement of the magnification and can be included in the maximum likelihood
inversion as a direct observable in order to break the 
 mass sheet degeneracy (see Sect. 3.2). Alternatively, it can also be used
to model the lens itself. In the case of  a singular isothermal sphere, the
magnification can be expressed as function of the velocity dispersion of
the lens, $\sigma$, and the radial distance $\theta=r/r_E$, where $r_E=4 \pi \sigma^2/c^2 \
D_{LS}/D_{OS}$:

\begin{equation}
\mu(r) = {4 \pi \sigma^2 \over c^2}\,{D_{LS} \over D_{OS}}\,{\theta \over
\theta-1} \ .
\end{equation}
Reconstruction of cluster mass distribution 
using magnification bias was initialy explored by Broadhurst et al
(1995), and has been used  by 
 Taylor et al (1998) in A1689, and by Fort, Mellier \&
Dantel-Fort (1997) in
 Cl0024+1654 (see also a generalization by Van Kampen 1998).  The 
masses found are consistent with those inferred from
 gravitational weak shear or from strong lensing.  \\
This magnification bias is an attractive alternative to the weak shear
 because it is only based on the 
galaxy counts and does not require outstanding seeing to measure 
ellipticities and orientations of galaxies.   However, it is
more sensitive to shot noise,
which unfortunately increases when the number density decreases in the 
depletion area.  Furthermore, it 
  also depends on the galaxy clustering of the background sources 
which can have large fluctuations from one cluster to another. Indeed,
in  the
weak lensing regime, assuming that $\kappa \approx \vert\gammag\vert$,
the ratio of the signal-to-noise ratios of the shear and the
 depletion, $R_{Sh/Dep}$, clearly favors the shear analysis:
\begin{equation}
R_{Sh/Dep}={\vert \gammag \vert \over \sigma_{\varepsilon}} {1 \over \kappa \vert (5 \alpha -2)\vert } \
\approx 3 
\end{equation}
for a dispersion of the intrinsic ellipticity distribution of the
galaxies, $\sigma_{\varepsilon} =0.3$, and $\alpha=0.2$. 
  Despite these limitations, this is a   
simple way to check the consistency of the mass reconstruction. Its 
 great merit is that it is not sensitive to systematic effects, like the 
 weak shear measurement which depends on
 the correction from the PSF  of the observed ellipticities.  
\section{LARGE-SCALE STRUCTURES AND COSMIC SHEAR}
The idea that mass condensations and the geometry of the Universe 
 can alter light bundles and distort the images of distant galaxies 
was emphasized by Christian \& Sachs 
(1966), and later by Gunn (1967) and Blandford \& Jaroszy\'nski (1981), 
who first gave a quantitative estimate of the 
amplitude of this effect. Kristian (1967) looked at this effect on 
photographic plates of  six clusters of galaxies using the Palomar
Telescope, but found nothing significant. Valdes, Tyson \& Jarvis (1983)
were the first who attempted to measure a coherent alignment of distant
galaxies generated by large scale structures. They used about 40,000 
randomly selected field galaxies with $J$  magnitudes between 
22.5 and 23.5, but, as Kristian, did not find  any conclusive signal. 
 These negative measurements were not definitely interpreted as 
 important cosmological constraints on the curvature and the mass
 distribution in our Universe, but rather as a result of technical limitations  
related to the poor image quality of the photographic data. 
  Indeed, the  recent  weak lensing analysis produced by a supercluster
candidate
 done by kaiser et al (1998) seems to show that large scale structures
 produce gravitational shear which is already detectable. 
Numerical simulations by 
 Schneider \& Weiss (1988) using point-mass models, 
 or Babul \& Lee (1991) using a 
smooth mass distribution, showed that both the ellipticity distribution 
 and the apparent luminosity function of distant galaxies could 
be modified, in particular if the fraction of small scale structures  
like clusters of galaxies is important (Webster 1985). Therefore 
 two different effects produced by the cosmological distribution 
 of structures in the Universe are expected: a change of the galaxy number 
count 
correlated with the mass distribution, namely a magnification bias; and
 a change of the ellipticity distribution, namely a shear pattern, 
 correlated with the mass distribution as well. Since the 
 expectation values strongly depend on the fraction of non-linear 
systems and the redshift distribution of the galaxies, 
 it is clear that the analysis of weak lensing effects by large-scale 
 structures  is an interesting test of cosmological scenarios. 
\subsection{Theoretical expectations}
The theoretical investigations of the effect of the large scale mass 
distribution on the distribution of ellipticity/orientation of distant
galaxies are somewhat simplified by the low density contrast of 
structures.  Beyond 10 Mpc scales, $\delta \rho /\rho \approx 1$ and
 linear  perturbation theory can be applied.  On these 
scales, lenses are no longer considered individually but they 
are now viewed as a random population  which has a 
cumulative lensing effect on the distant sources.  Blandford (1990), 
Blandford et al (1991) and Miralda-Escud\'e (1991) first investigated     
 the statistical distribution of distortions induced by large-scale
structures in an EdS universe.  They computed the two-point 
polarization (or shear) correlation function and 
 established how the rms value of the polarization depends on the power
spectrum of density fluctuations.  Kaiser (1992) extended 
 these works and showed how the angular power spectrum of the 
 distortion
is related to the three-dimension mass density power spectrum, 
without assumptions on the nature of fluctuation.  These works were 
 generalized   
 later to any arbitrary value of $\Omega$ by Villumsen (1996) and Bar-Kana
(1996). All these studies concluded that 
the expected  rms amplitude of the 
 distortion is of about  one percent, with a 
typical correlation length of a degree. Therefore 
it should be measurable with present-day telescopes. \\
These promising predictions convinced many groups to start
investigating more deeply how weak lensing maps obtained 
 from  wide field imaging surveys could constrain 
cosmological scenarios. To go into further details, it is 
 necessary to generalize 
 the previous works to any cosmology and to describe in detail 
 observables and physical quantities which could be valuable   
 to constrain cosmological models.  Indeed, the investigations 
 of weak lensing by large scale structures need theoretical and
statistical tools which are not different from those currently used 
 for catalogues of galaxies or CMB-maps.  In this respect, 
 the perturbation theory, which has already demonstrated to  
 work describe very well  the properties 
 of large scale structures (see Bouchet 1996 and references therein), 
looks  an ideal approach on such large scales, 
and the use of similar statistical estimators as for catalogues of
galaxies seems perfectly suited.   Bernardeau, Van Waerbeke \& Mellier 
 (1997) then used the perturbation theory in order to  explore 
the sensitivity of the second and
third moments of the gravitational convergence $\kappa$ (rather than the distortion 
 whose third moment should be zero),
 to  cosmological scenarios, and to cosmological parameters, including
$\lambda$-universes.  The small angle deviation approximation implies that 
 the distortion of the ray bundle can be computed on the unperturbed geodesic (Born
approximation). 
 In the linear regime, if lens-coupling is neglected (see Section 4.4.3), the cumulative 
 effect of structures along the light of sight generates a 
   convergence in the direction 
 $\thetag$  ,
\begin{equation}
\kappa(\thetag)={3 \over 2} \Omega_0 \ \int_0^{z_s} n(z_s) dz_s \ 
\int_0^{\chi(s)} {D_0(z,z_s) D_0(z) \over
D_0(z_s)} \delta(\chi,\thetag) \ (1+z(\chi)) d\chi \ , 
\end{equation}
where $\chi$ is the radial distance, $D_0$ the angular diameter
distance, $n(z_s)$ is the redshift
distribution of the sources, and 
\begin{equation}
\delta = \int \delta_k D_+(z) e^{i \boldk.\boldx} d^3 k 
\end{equation}
is the mass density contrast, which depends on the evolution of the
growing modes with redshift, $D_+(z)$. It is related to the power spectrum as usual
: $<\delta_{\boldk} \delta_{\boldk'}>=P(k) \delta_{Dirac}(\boldk+\boldk')$. 
 It is worth noting that $\kappa$
depends explicitly on $\Omega_0$ and not only $\delta$ because the
amplitude of the convergence depends on the projected mass, not only on the
projected mass density contrast. \\
The dependence of the angular power spectrum of
the distortion as a function of ($\Omega$,$\lambda$), of the power spectrum of
density fluctuations and of the redshift of sources has been  
investigated in detail   
 in the linear regime by Bernardeau et al (1997) and Kaiser (1998). 
 Bernardeau et al (1997) and Nakamura (1997) computed 
also  the dependence of
   the skewness of the convergence on cosmological parameters,
 arguing that it is the first moment which probes directly  non-linear
structures. From perturbation theory and assuming Gaussian 
 fluctuations, the variance,
 $ <\kappa(\thetag)^2>$, and the skewness, $s_3=<\kappa(\thetag)^3>/<\kappa(\thetag)^2>^2$, 
have the following
dependencies with the cosmological quantities: 
\begin{equation}
<\kappa(\theta)^2>^{1/2} \approx 10^{-2} \ \sigma_8 \ \Omega_0^{0.75} \ z_s^{0.8} \left({\theta \over  1^o}\right)^{-(n+2)/2}  \ ,
\end{equation}
and
\begin{equation}
s_3(\theta) \approx -42 \   \Omega_0^{-0.8} \ z_s^{-1.35}  \ ,
\end{equation}
for a fixed source redshift $z_s$, where $n$ is the spectral index of the power
spectrum of density fluctuations and $\sigma_8$ the normalization of the
power spectrum.
Hence, since the skewness does not depend on $\sigma_8$, the amplitude
of fluctuations and $\Omega_0$ can be recovered independently using 
$<\kappa(\theta)^2>$ and $s_3$. \\ 
\begin{figure}
\centerline{
\psdraft{figure=mellfig06.eps,width=7cm}}
\caption{\label{jainseljak.ps} {\it Ratio of the amplitude of the polarization 
 predicted by the non-linear and the linear evolution of the power spectrum as function
of angular scale (from Jain \& Seljak 1997). The normalization is $\sigma_8=1$. The
plots shows the expectations for three cosmologies: $\Omega=1$ (solid line),
$\Omega=0.3$ (dashed line) and $\Omega=0.3$, $\lambda=0.7$ (dotted line). The
difference between the two regimes becomes significant below the 10' scale.} 
}
\end{figure}
Jain \& Seljak (1997), generalizing the early work by Miralda-Escud\'e
(1991), 
 have analyzed the effects of non-linear evolution on 
$<\kappa(\theta)^2>$ and $s_3$ using the fully non-linear evolution 
of the power spectrum (Peacock \& Dodds 1996).  They found formal 
 relations similar to 
those found by Bernardeau et al. However, on scale below 
10 arc minutes, 
$<\kappa(\theta)^2>$ increases more steeply than 
  the theoretical expectations of the linear theory and  
 is 2 or 3 times higher on scales below 10 arcminutes. 
These predictions are strengthened by  
 numerical simulations (Jain, Seljak \& White 1998). Therefore, 
 a shear
amplitude of about 2-5\% is predicted on these scales which should be
observed easily with ground-based telescopes (Figure \ref{jainseljak.ps}). 
 Schneider et al (1998a) 
  recently claimed that they have detected this small-scale {\it cosmic-shear} signal. \\
The previous studies are based on the measurements of ellipticities 
of individual galaxies in order to recover the stretching produced by
linear and non-linear structures.  As for the mass reconstruction of
clusters, it demands high quality images and an accurate correction
of systematics down to a percent level.  An alternative to this 
 strategy has been investigated by Villumsen (1996) who looked at the
effect of the magnification bias on the two-point galaxy correlation
function. Since the magnification may change the galaxy number density
as function of the slope of the galaxy number counts, it also modifies
the apparent clustering of the galaxies is a similar way. From Eq. (23)
the two-point correlation function averaged over the directions
$\theta$ is changed by the magnification of the sources
 and, in the weak lensing regime, its contribution
  writes (Kaiser 1992, Villumsen 1996, Moessner \& Jain 1998):
\begin{equation}
\omega(\theta)={\left<\left[N_0(m)(5\alpha-2)\kappa(\theta+\theta')\right]
\left[N_0(m)(5\alpha-2)\kappa(\theta)\right]\right>\over N_0(m)^2} \ ,
\end{equation}
that is 
\begin{equation}
\omega(\theta)=(5\alpha-2)^2 \ \left<\kappa(\theta+\theta')\kappa(\theta)\right> \ .
\end{equation}
The galaxy two-point correlation function is therefore sensitive to the correlation 
function of the convergence and to the slope of galaxy counts. 
Provided that the unlensed two-point correlation function is known, it is
then possible to compute the local correlation function of the convergence 
from the local two-point correlation function of the galaxies. \\
Details investigations of the capability of this technique have been
discussed by Moessner, Jain \& Villumsen (1998a) who looked at the 
 effect of non-linear clustering on small scales and for different   
cosmologies. They raised the point that the correlation function 
can be also affected by the evolution of galaxies which modifies also 
the two-point correlation function of galaxies, but in an  unknown
way. Moessner \& Jain (1998)  proposed to disentangle these two effects
 by using the  cross-correlation of
two galaxy samples having different redshift distributions which do 
 not  overlap. This minimizes the effect of intrinsic galaxy clustering. 
However it requires the knowledge of the biasing which can also depend
on the redshift. Therefore, the magnification bias method needs 
auxiliary input which can constrain the biasing independently.
\subsection{Measuring the biasing}
Since gravitational lensing is directly sensitive to the total mass
responsible for the deflection,  it provides a potentially important tool
for measuring the biasing factor, 
as it has been demonstrated by the recent
 weak lensing  analysis of the Supercluster MS0302+17 (Kaiser et al
1998).   In particular, a well-known effect of
 the magnification bias is the generation  
of  correlations between foreground and background luminous systems 
 observed in catalogues. The matter associated with the foreground 
systems can amplify the flux received from background objects, which 
  results in an apparent correlation between the number density or the 
 luminosity of backgound objects and the number density of foregrounds.
 The correlation has been first detected by Fugman (1990) and confirmed 
 latter by Bartelmann \& Schneider (1993a,b; 1994) who found 
correlations of the galaxy number density with 
 the radio-sources of the 1-Jy catalogues, 
 the IRAS catalog, and  the X-ray galaxies.  
 Further independent analyses of such associations showed also evidences of
magnification bias  (Bartelmann, Schneider \& Hasinger 1994; Rodrigues-Williams \& Hogan 1994; 
Seitz \& Schneider 1995b; Wu \& Han 1995; 
Ben\'{\i}tez \& Mart\'{i}nez-Gonz\'alez 1997; Williams \& Irwin 1998).  The 
 statistical basis of these associations is surprisingly  robust and 
 difficult to explain physically without invoking lensing
effects.  \\
Bartelmann et al (1994) argued 
 that the correlation can be interpreted  
 has a magnification bias only if it is produced by condensations 
 of matter like clusters of galaxies (Bartelmann \& Schneider 1993a). 
 In order to check this hypothesis, Fort et al (1996) have 
attempted to detect weak shear around some selected bright quasars 
 which could be good lensing candidates. They found
strong evidence of weak shear  around half of their sample with,   
in addition, a clear detection of galaxy overdensities in the 
 neighborhood of each quasar. The Fort et al
sample has been re-analyzed recently
by Schneider et al (1998a) and the detection has been strengthened on 
 a more reliable statistical basis, with a firm confirmation 
 for one of the quasars from the  HST images obtained by Bower \& Smail
(1997).  In addition, some other 
 observations have also detected gravitational shear around 
 bright radiosources:
 Bonnet et al (1993) found a shear signal around the double
imaged quasar Q2345+007, which was later confirmed to be associated with 
 a distant cluster (Mellier et al 1994; Fischer et al 1994; Pell\'o et al 1996).
 A similar detection of a  cluster has recently been reported 
 around the Cloverleaf (Kneib et al 1998) and   
 around 3C324, which also clearly shows a shear pattern from
 the HST images (Smail \& Dickinson 1995).
   The sample is nevertheless too small to provide a 
significant direct evidence  that 
magnification bias is detected in quasar catalogues. Indeed, it would be important to 
pursue this
 program using a large sample of bright quasars or radiosources. 
The field of view does need to be large, so the HST with the Advanced
Camera could be a perfect instrument for this project. The impact of
such magnification bias could be important for our understanding of the 
evolution scheme of quasars and galaxies, since it changes the apparent
luminosity functions of these samples (Schneider 1992, Zhu \& Wu 1997).  \\
If these correlations are due to magnification bias, a quantitative 
value of the biasing can  in principle be estimated, for instance 
 from  the angular
foreground-background correlation function (Kaiser 1992), where
foreground could be galaxies or dark matter. Bartelmann (1995a) expressed
the angular quasar-galaxy correlation, $\xi_{QG}$,
  as function of the biasing factor, $b$, 
and the magnification-mass density contrast cross-correlation 
function, $\xi_{\mu\delta}$, in the weak lensing regime: 
\begin{equation}
\xi_{QG}(\thetag)=(2.5 \alpha-1) \ b \ \xi_{\mu\delta}(\thetag)
\end{equation}
where $\alpha$ is the slope of the background galaxy counts. 
 However, since $\xi_{\mu\delta}$ depends on the power spectrum of the 
projected
mass density contrast, the determination of the biasing factor 
 is possible only if one obtains  independently the amplitude of the power 
spectrum. Generalizations   
 to non-linear evolution of the power spectrum and to 
 any cosmology were explored by Dolag
\& Bartelmann (1997) and Sanz, Mart\'{\i}nez-Gonz\'alez \& Ben\'{\i}tez (1997). As 
expected, the non-linear condensations increase the correlation on small
scales by a very large amount, but it still strongly depends on the 
 amplitude of the power spectrum. \\
With the observation of weak lensing induced by large scale structures,
it becomes possible to observe directly the correlation between 
the background galaxies and the projected mass density of foregroung structures 
 instead of using the light distribution emitted by the foreground galaxies. 
 Following the earlier development by Kaiser (1992),
 Schneider (1998)
 has computed the cross-correlation between the projected mass, $M$, and the galaxy 
number density of foreground galaxies, $N$,  on a given scale $\theta$, $<MN>_{\theta}$,
 as function of the biasing
factor of the foreground structures, $b$. Like previous 
 studies,  it is simply proportional 
to $b$, but it also depends on 
$\sigma_8$ and the slope of the power spectrum, and as such it is not 
trivial to estimate it without ambiguity. A more detailed investigation of the
 interest of $<MN>_{\theta}$ has been done by Van Waerbeke
(1998a) who computed   
 the ratio, $R(\theta)$ of the density-shear
correlation over the two-point galaxy correlation function 
  for a narrow redshift distribution of foregrounds 
 and a narrow range in scale (Van Waerbeke 1998a,b):
\begin{equation}
R_{\theta}={3 \over 2}{\Omega \over b}{g(w_f) f_K(w_f)N_f(w_f) \over 
a(w_f) \ \ \int N^2_f(w) dw} \ ,
\end{equation} 
where $a$ is the expansion factor,  $w_f$ is the comoving distance of the foreground, 
$f_K$ the comoving angular diameter
distance, $N_f$ the redshift distribution of the foreground galaxies,
and 
\begin{equation}
g(w)=\int_{w}^{w_{f}} N_b(w') {f_K(w'-w) \over f_K(w')} 
\end{equation}
$N_b(w')$ being the redshift distribution of the background galaxies. 
 Using this ratio, he investigated 
 the scale dependence of the biasing,  including  the non-linear
spectrum and different cosmologies, and  discussed 
 how it can be used to analyze the evolution of the biasing with redshift,   
   if two different
populations of foregrounds are observed. 
 The ratio  $R_{\theta}/R_{\theta'}$ 
  permits to compare the biasing on two 
different scales. This quantity does not depend 
 on $\Omega$, neither on the power  spectrum nor on the 
 smoothing scale, so it is a direct estimate of the evolution of biasing
with scale.  Van Waerbeke  predicts that 
 a variation of 20\% of the bias on scales between 1' and 10' 
 will be detectable on a survey covering 25 square degrees. 
This ratio is therefore a promising estimator of
the scale dependence of the biasing.  
\\
The analysis discussed above indicates the potential 
 interest of lensing to study   
 the evolution of bias with  scale and redshift of 
foreground, using directly the correlation between the ellipticity distribution 
 of background sources and the projected mass density inferred from 
 mass reconstruction.  It thus permits an accurate and direct 
 study of the biasing and its
evolution with look-back time and possibly with 
 scale.  However, it is worth  stressing
  that Van Waerbeke (like Bartelmann 1995a and others) assumed a 
linear biasing. In the future it 
will be important to explore  in details the generalization to 
 a non-linear bias (Van Waerbeke, private communication) as 
described by Dekel \& Lahav (1998).
\subsection{Strategies for weak lensing surveys}
The rather optimistic predictions from theoretical work reported 
above are convincing enough to investigate in great detail how 
weak lensing surveys can be designed in order to maximize the 
 signal to noise ratio of statistical and physical quantities, as well 
 as to minimize the time spent for the survey. The definition of a 
best strategy addresses many issues: size, shape, topology of the survey,
best statistical estimators, optimal analysis of the catalogues and 
 best extraction of the signal from the raw data. This last point 
 (measurement of ellipticities, correction of the PSF)  has already been  
discussed in previous sections, so I will focus on the other aspects.
 Indeed, weak lensing surveys have rather generic constraints which 
are common to any survey in cosmology (see Szapudi \& Colombi 1996,
Kaiser 1998, Colombi, Szapudi \& Szalay 1998 and references therein).  
The main  sources of noise 
are the cosmic variance,  the intrinsic ellipticity distribution of the 
lensed galaxies and possibly the noise propagation during the mass 
reconstruction from lensing inversion. In addition, at such low shear 
level, the correction of the PSF, as well as the 
  removals of systematics coming from optical and atmospheric degradation 
of the images, or from the method used to measure the shear 
from galaxy ellipticity, are crucial steps. \\
The impact
of the size, of the shape and the deepness of weak lensing surveys has been 
addressed by Blandford et al (1991) and Kaiser (1992) 
and investigated later in more detail by Kaiser (1998), Kamionkowski et al
(1997), Van Waerbeke et al (1998),  and Jain et al (1998).  Depending 
 on the scientific goals and the nature of the data, medium-size deep,  
very-wide shallow  and very-small ultra-deep surveys have been 
 proposed.\\
The most detailed investigation done so far has been 
conducted by Van Waerbeke et al (1998). From a sample of 60 simulations
per set of parameters they computed the expected signal to noise ratio 
 of the variance and the skewness of the convergence, and  
 attempted to reconstruct the projected mass density from the lensing 
signal in order to recover the 
 projected power spectrum for each model 
on 5$^o\times$5$^o$ and 10$^o\times$10$^o$ noisy maps.   They included  various
cosmologies, mass density power spectra, sampling strategies and 
redshifts of sources. This work is a preparation of
the sub-arcsecond seeing survey which should be conducted at CFHT 
with the new wide field MEGACAM camera (which will cover one 
 square   degree, Boulade et al 1998).  These simulations show that 
 the projected mass density
maps can be recovered with an impressive accuracy and demonstrate 
that such a survey will be able to recover the projected power
spectrum
of mass density fluctuations, from 2.5 arcminutes up to 2 degrees with a
signal to noise ranging from 10 to 3.
The power spectrum on small scale is dominated by the shot noise 
 due to the intrinsic ellipticity distribution 
of the background galaxies, but it follows the statistics predicted by
Kaiser (1998) which permits to remove it easily.  
The best strategy suggested by these simulations is 
 a shallow survey 
 with a typical galaxy number density of 30 arcmin$^{-2}$ 
over 10$\times$10 square degrees. According to 
 Van Waerbeke et al (1998) it will permit to separate  $\Omega=0.3$ 
 from   $\Omega=1$ universes  at a 6 $\sigma$ confidence level (Figure
\ref{omegalambda1.ps}).  
Kaiser (1998) recommends a wide field 
as well, but suggests a sparse sampling on very large scale rather than
a compact topology. This alternative has not been investigated yet 
using simulations. In particular, it would be useful to have
quantitative estimates of the scale beyond which the survey should switch 
from a compact to a sparse sampling of the sky.
\\
\begin{figure}
\centerline{
\psdraft{figure=mellfig07.eps,width=8cm}}
\caption{\label{omegalambda1.ps} {\it Histograms of the values of the
skewness of the convergence for an $\Omega=1$ (right) and $\Omega=0.3$ (left) universes and
for a 5$\times$5 (thick lines) and 10$\times$10 (thin lines) square
degrees survey. This simulates the possible surveys to be done 
 with MEGACAM at CFHT (from
Van Waerbeke et al 1998).} 
}
\end{figure}
Alternatively, Stebbins (1996) and Kamionkowski
et al (1997) have explored the possibility to use shallower surveys, 
 which sample the sky with only  few galaxies per arcminutes, 
 but  cover about half the sky.   Contrary to the MEGACAM-like
  surveys 
 which aim at mapping the shear field and to build a map of the 
projected mass density, these shallow surveys only aim at measuring the
correlation of ellipticities on very large scales. 
  Stebbins (1996, 1999) used the linear theory to compute the 
angular power spectrum 
 of the shear inferred from a  tensor spherical harmonic expansion 
 of the shear pattern of an all-sky survey.  
 He argues that  the SDSS could provide reliable information on the projected 
angular power 
spectrum of the shear. The expected signal is very small, because 
most of the galaxies will be at redshift lower than 0.2.  The seeing of
the site and the sampling of the images could have minor impact of the
signal because these nearby galaxies should be much larger than the
seeing disk.  However, a simulation of the systematics which include 
the quality of the telescope and the instruments, as well as 
 an estimate of the possible systematics produced by the drift-scanning 
is now necessary in 
order to have a clear quantitative estimate of the expected signal to
noise ratio.  \\
The use of the VLA-FIRST radio survey for weak lensing proposed by 
 Kamionkowski et al (1997) looks promizing. FIRST 
 covers about the same area 
as the SDSS but the radiosource sample has a much broader redshift
distribution and a median redshift beyond $z=0.2$. However, 
with  less than 100 sources per square-degree, the sampling 
 is rather poor. Kamionkowski et al predict 
an rms ellipticity of about 3\% on 6 arcminutes, 
 1\% on 20 arcminutes and on one degree scale, with a signal-to-noise ratio
larger than 5, 
in good agreement with Bernardeau et al. (1997) and Jain \& Seljak
(1997). Preliminary results
from this survey should come out quickly.
\\
On small scales, first attempts for measuring cosmic shear have 
already been made by several groups. Mould et al. (1994) used a very deep
image obtained at the Palomar Telescope to measure cosmic shear 
 on scale of 5 arcminutes.  They do not 
find a significant signal (see however Villumsen 1995, unpublished results) 
 and argued that the cosmological signal 
 of their field is below 4\%. A similar attempt has been 
done by  Fahlman et al. (1994) on a 15 arcmin field. The 
images were obtained at CFHT in subarcsecond
seeing conditions ($\approx$ 0.5", against 0.9" for Mould et al), but significantly less deep than the
Mould et al. observations. 
 They did not detect any significant signal, though, due to the excellent
seeing, they expected to be 3 times more sensitive.  On the other hand,
 the detections of weak lensing signal on about 2 arcminute 
scale around radiosources and quasars  (see previous section on biasing)
 show that cosmic shear on small scales is 
already detectable even for shear amplitudes of 2\%. 
Schneider et al (1998a) argued that the shear 
detected around radiosources is cosmic shear, though the sample is probably 
biased toward non-linear structures because the fields are preferentially
selected around bright quasars. This is probable because the
motivation of the Fort et al (1996) observations (from which the
Fort et al  and Schneider et al sample is based) was 
the verification of the Bartelmann et al (1994) hypothesis that
the quasar-galaxy association results from a magnification bias  of
 bright quasar samples by large scale structures.  
  Therefore, if 
 the magnification bias really works then the
 cosmological significance of the Schneider et al (1998) interpretation is
difficult to quantify. Conversely, if the quasar-galaxy associations are
not produced by such magnification, then Schneider et al measured the
cosmic shear for the first time.  \\
\begin{figure}
\centerline{
\psdraft{figure=mellfig08.eps,width=18cm}}
\caption{\label{shearkappacarte.ps} {\it Simulation of mass maps construction from 
a wide field weak lensing  survey. The left panel is the original simulated projected mass
density of large scale structures. The field covers 25 square degrees. The middle 
panel is the reconstructed mass map using the algorithm described in Van Waerbeke et al
(1998). The noise introduced in the simulated map is due to the intrinsinc ellipticity
distribution of the lensed galaxies. The similarity with the original
mass map is striking. The right panel is the shear map overlayed to the projected
mass map. The length and the orientation of each line indicate  the amplitude 
 and the orientation the shear. 
} 
}
\end{figure}
Several low-angular scale surveys are underway in order to provide 
much better statistics. They have been summarized by Seitz 
(1999). A promising  strategy consists in using the parallel STIS
observations of HST, as randomly selected fields. For each of those
outstanding images, it is possible to measure the average
distortion with a very high confidence level. Preliminary 
analyses by Seitz (1999) show that the PSF is incredibly stable
 and that the first observations  lead to an rms ellipticity of 3\% 
 on one arcminute scale, which 
is very encouraging. However, for the moment it is difficult to interpret 
 the rms shear in terms of constraints for cosmological 
models because the STIS parallel observations are done 
 without filter, so there is no color informations on the lensed galaxies 
 to constrain their average redshift. \\
In addition to the definition of the survey strategy, the scientific
return is sensitive to the technique used to analyze the catalogues. 
In this respect, the best procedure depends on the noise properties 
on the survey. In theory, if the noise follows a Gaussian 
statistics, the Wiener filtering should provide a minimum variance 
estimator of cosmological quantities (Seljak 1997a). However, it is not
obvious that this is the best approach, in particular on small scales 
where non-linear features deviate significantly from gaussianity.
 For instance, Kruse \& Schneider (1998) have explored a strategy 
for an optimal extraction of the high density peaks present in surveys.  
It uses the aperture mass densitometry for cosmic 
 shear measurement proposed by Schneider et al
(1997) which permits to detect peaks of the projected mass density 
 as function of the tangential shear, $\gammag_t$ (Kaiser
1995). This simple and rather robust approach  which focuses 
 on the most contrasted systems should provide in a simple way   
some constraints of the number density of high peaks, and therefore 
on the cosmological scenarios. Schneider et al (1997) argued that 
 the use of a compensated filter is an optimal procedure 
 to measure $\kappa$ from $\gammag$ without mass reconstruction because in 
contrast to the top hat filter, it selects a sharp scale range which 
 cancels the additional noise produced by the adjacent wavenumbers.  
 The merit of the two filterings has also been investigated in detail in the 
simulations of Van Waerbeke et al (1998). They found that the best
choice is unclear: the compensated filter is  better for the variance, but 
is worse for the skewness. \\
\subsection{Critical issues}
The critical issues already discussed for the weak lensing and mass
reconstruction of clusters of galaxies are valid and even more critical
for large scale structures. This will not be re-discussed here, but we
 should mention that systematic effects are a real concern which could be an
ultimate limitation of the weak lensing capabilities, at least on ground
-based telescopes.   
\subsubsection{Redshift of sources}
Eqs.(27) and (28) show a strong dependence of the variance and the
skewness of the convergence on the redshift distribution of the 
lensed sources.  As far as weak lensing is concerned, from 
 the investigations of the effect of  redshift distribution of sources 
 on cluster mass reconstruction, it seems that only the averaged 
redshift of the galaxies and the width of their distribution 
 are needed (Seitz \& Schneider 1997), even with a bad 
precision (say, $\Delta z_s \approx 0.5$). This requirement is 
therefore not severe and can be 
obtained using photometric redshifts. The  work  by
Connolly et al. (1995) demonstrates that galaxies brighter than $I=22.5$
can easily be calibrated using spectroscopic redshifts and 
 that photometric
information in 4 different filters constrain the redshift of these
galaxies with an accuracy of about $\Delta z_s \approx 0.4$.  The future 
surveys with 10-meters class telescopes will calibrate 
photometric redshift up to $I=23.5$, and even fainter by using 
Lyman-break galaxies (Steidel et al 1998) and near-infrared photometry.  
This is deep enough for shallow wide field surveys so one can be 
reasonably confident that the redshift distribution of the lensed 
galaxies will not be a major problem. Conversely, this is another
argument in favor of shallow rather than deep surveys which would 
reach limiting magnitudes beyond the capabilities of spectrographs. 
Indeed, as shown by Van Waerbeke et al (1998) the signal to noise ratio
of the variance and the skewness   
does not strongly depend on the redshift of the sources, so it is useless
 to reach very faint magnitudes.
\subsubsection{Source clustering}
Due to the intrinsic clustering of the galaxies
 the redshift distribution can be broad enough to mix together the
population of lensed galaxies and the galaxies associated with the
lensing structures. In particular an extended massive structure at high 
redshift
can play simultaneously the role of a lens and a reservoir of lensed
galaxies. The average redshift distribution of the sources can therefore
be biased by the galaxies located within the massive structure,  
which can bias as well the estimated value of the convergence in a similar way.
Indeed, the variance of the convergence is not affected by this
clustering (Bernardeau 1998a). However, the skewness is much more 
affected, mainly because
the overlapping acts exactly as a non-linear evolution of the
projected density.  Bernardeau (1998a) shows that most these effects are
negligible on scales beyond 10 arcminutes. It would  also be interesting to 
investigate more deeply how the source clustering may contribute to 
spurious signal on small scales.\\
The apparent change of the two-point correlation function by
magnification bias can also change the local redshift distribution 
of lensed sources. This effect, though mentioned by Bernardeau (1998a),
has not been investigated in detail.
\subsubsection{Lens coupling}
When ray bundles cross, by accident, two
lenses, the cumulative convergence is given by the product 
of the magnification matrix, and not simply the sum of the 
two convergences. The resulting convergence contains additional coupling terms 
 that must be estimated. 
 Fortunately,  in the weak lensing regime, 
 this coupling appears to be  negligible. It does not
change the value of the variance, and the skewness is 
 only weakly modified (Bernardeau et al
1997; Schneider et al 1997).   
\subsubsection{Validity of the Born approximation}
The effects of mass density fluctuations due to large scale structures 
  on the deformation of the ray bundles are computed assuming that the Born
approximation is valid, that is the deformation can be computed
along the unperturbed geodesic. In the case of linear perturbations, 
 this assumption is valid at the lowest order.  As discussed by Bernardeau et al
 (1997), the correction on the skewness should be at the percent level.
However, this is less obvious once lens couplings are taken into account. 
  The validity of the Born approximation has not been tested in detail, so far.  
  This certainly deserves inspections using high-resolution numerical
simulations. The simulations done by Jain \& al (1998) or
Wambsganss, Cen \& Ostriker (1998) could provide valuable informations on this
issue.
\subsubsection{Intrinsic correlated polarization of galaxies}
It is possible that the intrinsic orientations of galaxies  
 are not randomly distributed but have a 
coherent alignment correlated to the geometry of large
scale structures in which they are embedded. If it is so, the coherent 
alignment produced 
by weak lensing will be contaminated by the intrinsic 
 alignment of the galaxies and a mass reconstruction 
 based on the shear pattern will be impossible. Such alignments 
could appear during the formation of large scale structures 
 or could result from dynamical evolution of galaxies
within deep potential wells, like superclusters or clusters of galaxies
(see Coutts 1996; Garrido et al 1993 and references therein).
However, the most recent numerical simulations do not show such   
correlations.
 Many attempts have been done to search for signatures of these 
 intrinsic coherent patterns. So far, no convincing observations 
 on nearby structures have
demonstrated that there are large-scale coherent alignments.
 This possibility has to be investigated deeply, in particular in nearby
large scale structures where a coherent alignment from gravitational
lensing effect is negligible. It would be valuable to have more
 quantitative estimates of possible trends for alignments  
 from future very high-resolution numerical simulations of structure formation.
\section{GALAXY-GALAXY LENSING IN FIELD GALAXIES}
Evidence that galaxies have dark halos come from the kinematical and
dynamical studies of galaxies.  However, the geometry of the halos, and
the amount and distribution of dark matter are unknown and
in practice difficult to probe.  Gravitational lensing could 
provide valuable insight in this field: since it  works on
all scales, in principle the halos of galactic dark matter should 
be observed from their gravitational lensing effects on 
 background galaxies.  The first Einstein rings and the other galaxy-scale lensing 
 candidates   
have provided unique opportunities to measure  
 the mass-to-light ratios and to probe the mass profiles
of a few galaxies (Kochanek 1991). In the case of rings, the mass 
 of the lensing galaxies can be very well constrained (see for instance Kochanek 1995), 
so  the properties of the
halos inferred from modeling are reliable.  However, Einstein 
rings are rare lensing events, so the sample 
 is still very small.\\
A more promising approach consists in a statistical study of the 
  deformation of distant galaxies by foreground galactic halos.  
 The galaxy-galaxy lensing analysis uses the correlation  
 between the position of foreground galaxies and the position-alignment
 of their angular-nearest neighbors among the background population. 
If the alignment is assumed to be produced by 
 the gravitational shear of the foreground halos, then it is 
possible to probe their mass, if the redshift distributions of the
foregrounds and the backgrounds are known. A statistical analysis is 
then possible, if one assumes that all the foreground galaxies have similar halos,
 which can be scaled using the Tully-Fisher relation and the 
 photometric data (galaxy luminosity).  The 
 expected gravitational distortion is very weak: for foregrounds at
redshift $<z_l>=0.1$, backgrounds at $<z_s>=0.5$, and typical halos
with velocity dispersion of 
 200 kms$^{-1}$ and  radius of 100 kpc, $\vert \gammag \vert \approx
$1\% at about 20 kpc from the center. But if the observations go 
 to very faint magnitudes
there is a huge number of background lensed galaxies, so that the 
weakness of the signal is compensated by the large statistics. It is 
worth pointing out that the signal-to-noise ratio depends on the number of galaxy
pairs, and so either very wide field shallow surveys or ultra-deep 
 imaging can be used for the statistics (though they will not probe 
similar angular scales). \\
A very
first attempt was done by Tyson et al (1984) who used 
 about 50,000 background and 11,000 foreground galaxies obtained from   
photographic plates.  The 3 $\sigma$ upper limit of the circular
velocity they found was 160 km.sec$^{-1}$, with a maximum cutoff 
radius below 50 $h_{100}^{-1}$ kpc. These values are significantly smaller
than theoretical expectations from rotation curves and dynamical
analyses of galaxies.  However, despite  a careful examination of possible 
systematics (Tyson et al 1984, Tyson 1985), there are two 
limitations to Tyson et al's pioneering work.  First, as emphasized by
Kovner \& Milgrom (1987), the assumption that background galaxies are at
infinite distance has considerable impact on the constraints on the 
circular velocity and the cutoff radius.  If  
 one includes a corrective factor which takes into account the distances
of the sources,  the upper limit for the circular velocity
is considerably higher (330 kms$^{-1}$ for a $L_{*}$ galaxy) 
and no constraints can be put on the cutoff radius (Kovner \& Milgrom
1987).  Second, the image
quality of the photographic plate is  bad and  
 may also affect the measurement of weak distortions.\\
\begin{figure}
\centerline{
\psdraft{figure=mellfig09.eps,width=10cm}}
\caption{\label{galgalbrainerd.ps} {\it Angular variation of polarization produced by
weak lensing of foreground galaxies on the background (lensed) sources in the Brainerd 
 et al (1996) sample. The lines show theoretical expectations for three models 
 of halos having different velocity dispersion and scales.}  
}
\end{figure}
The first attempt to use deep CCD subarcsecond images has been 
done by Brainerd, Blandford \& Smail (1996) using about 5,000 galaxies.
The distortion was compared with simulations, based on 
analytical profiles for the dark matter halos, and the Tully-Fisher
relation, in order to relate mass models to observations.  After careful
investigations of systematics, they detected a significant polarization 
of about 1\% , averaged over separation between 5" and 34" (see Figure
\ref{galgalbrainerd.ps}).  They
concluded that halos smaller that 10$h^{-1}$ kpc are excluded at a
2$\sigma$ level, but the data are compatible with halos of size
larger that 100$h^{-1}$ kpc and circular velocities of 200 kms$^{-1}$.
 Similar works using 23,000 galaxies  has been done recently 
 using very deep images obtained with MOCAM at CFHT (Cuillandre et al 1996) 
 with seeing below 0.7". They found
remarkably similar results as Brainerd et al for the polarization and 
its evolution with radius (Erben et al, in preparation).\\
The HST data look perfectly suited for this kind of program which
demands high image quality and the observation of many field galaxies.
Griffiths, Casertano \& Ratnatunga (1996) used the Medium Deep Survey 
 (MDS) and measured the 
distortion produced by foreground elliptical and spiral galaxies. They
found similar results as Brainerd et al (1996) but with a more significant
signal for foreground elliptical than spiral galaxies. The comparison 
 with shear signals expected from various analytical models seems  to 
 rule out 
 de Vaucouleur's law as mass density profile of ellipticals. Ebbels, Kneib \& Ellis 
 (in preparation) are now extending the MDS work to a larger sample, trying to
simulate more carefully the selection effects.   
  Dell'Antonio \& Tyson (1996) and Hudson et al (1998)
analyzed the galaxy-galaxy lensing signal in the HDF. As
compared with the ground-based images or the MDS, the
field is small but the depth permits to use many background galaxies 
 even on scale smaller than 5 arcseconds.  Furthermore, the UBRI 
 data of the HDF permit to infer
 accurate photometric redshifts for the complete sample of galaxies.
 Dell'Antonio \& Tyson compared the lensing signal with predictions from
an analytical model for the halo.
They found a significant distortion of about 7\% at 2" from the
halo center which corresponds to halos with typical circular velocities
of  less than 200 km.sec$^{-1}$.  The results 
obtained by Hudson et al (1998) are consistent with those 
of Dell'Antonio \& Tyson (1996) and Brainerd et al (1996).
 However, their maximum-likelihood analysis permits to take 
 into account more accurately the collective effects of large-sized
halos (Schneider \& Rix  1997). In contrast to previous studies,
 they made careful corrections of images from the PSF and scaled the 
magnitude inferred from the analytical models of the halos using the 
Tully-Fischer relation. \\
Galaxy-galaxy lensing is potentially a very valuable tool
 to study the dynamics of galaxies, complementary of standards 
 methods using photometry and spectroscopic data. Since 
the foreground galaxies have an average redshift of about 0.1, 
  it offers also 
an opportunity to look at the dynamical evolution of galaxies by
 comparing local galaxies at redshift zero to intermediate-redshift 
 galaxies (Hudson et al 1998). However, there are still some limitations
 due to the rather small number of galaxies used in each sample.  
 As shown from the simulations by Schneider \& Rix, it is rather easy to
constrain the velocity dispersion of the halos, but more difficult 
 to put  limits of their physical scale (see Figure \ref{rixgalgal2.ps}).  
There are also uncertainties
coming from the models of halos (which are assumed spherical) and the 
 additional noise produced by cosmic shear which can contaminate the
galaxy-galaxy signal. Though these issues should be analyzed in more
detail in the future, dramatic changes in the results are not expected
 (Schneider \& Rix 1997). In particular, the weak lensing induced by
large-scale structures should  indeed be cancelled by the averaging
procedure of  galaxy-galaxy lensing.
\begin{figure}
\centerline{
\psdraft{figure=mellfig10.eps,width=7.cm}}
\caption{\label{rixgalgal2.ps} {\it Simulations of galaxy-galaxy lensing
done by Schneider \& Rix (1997) for four samples of galaxies: 795 (top left),
1165 (top right), 2169 (bottom left) and 3137 (bottom right).  The {\rm O} symbol 
indicates the input parameters  of the model and the {\rm X} show the
maximum of the likelihood function.  The likelihood isocontours shows that the velocity 
dispersion is recovered easily, in contrast to the scale which requires many more 
galaxies.}
}
\end{figure}
\section{GRAVITATIONAL TELESCOPE AND HIGH-Z UNIVERSE}
\subsection{Redshift distribution of galaxies beyond B=25}
 Gravitational lensing magnifies part of  the distant universe and  
permits to explore the redshift distribution of 
 faint galaxies as well as the morphology and the contents of 
 very distant galaxies.  As it was
discussed previously, information on the distances of the sources are 
relevant  for the weak lensing inversion, since the mass reconstruction  uses
a grid of  faint distant sources whose redshift distribution is 
 basically unknown. 
 In particular, this hampers the mass estimates 
of high-redshift 
 lensing clusters which are very sensitive to the redshifts of 
the background sources (Luppino \& Kaiser 1997).   Unfortunately, 
 beyond $B=25$, even giant optical telescopes are too small for spectroscopy and
the redshift of a complete sample of $B>25$ galaxies cannot
be secured in a reasonable amount of observing time. The possibility of
using photometric
redshifts is very promising, but observations
as well as tests of the reliability of the method for the faintest galaxies are 
still underway 
 and deserve careful control. Moreover, since it is hopeless
  to calibrate the photometric redshifts of the faint samples
 with spectroscopic data, a cross-check of the predictions of
photometric redshift and ``lensing-redshift'' is important.\\
\subsubsection{Spectroscopic surveys of arclets}
Spectroscopic redshifts of arc(let)s is a long and difficult task 
but which is definitely indispensable for lensing studies. They permit  
to compute the angular distances $D_{OL},
D_{LS}$ and $D_{OS}$ and therefore to get the absolute scaling of the
projected mass density. These redshifts
 probe also directly the positions of critical lines
 which eventually constrain the local mass distribution for some 
 detailed models (Kneib et al
1993, 1996; Natarajan et al 1998; Kneib et al, in preparation). The development of the 
lensing inversion
technique (see next section) also demands   spectroscopic 
confirmations of its predictions in order to demonstrate that this is 
 a reliable and efficient method.  Last but not least, it
is in principle also possible to obtain information on the cosmological 
parameters if one
could have enough redshifts to constrain both the mass distribution 
 of the lens and the geometry of the Universe.\\
Spectroscopic surveys of the "brightest"
arclets in many clusters  are progressing well. Some
 extremely distant galaxies have been discovered, 
like the arc(let)s in Cl1358+6245 at $z=4.92$ 
(Franx et al 1997, see Figure \ref{cl1358.ps}), in A1689 at $z=4.88$ 
(Frye \& Broadhurst, private 
 communication), 
 in A2390 at $z=4.04$ (Frye \& Broadhurst 1998),  in 
 Cl0939+4713, where three $z>3$ arcs have been 
 detected (Trager et al 1997), or the  hyperluminous
galaxy in A370 at $z=2.8$ (Ivison et al 1998).  Extensive 
 spectroscopic follow-up is also 
 underway  in A2390 (B\'ezecourt \& Soucail 1996, Frye et al 1998)
and in A2218 (Ebbels et al 1996, 1998). The spectroscopic
observation by Frye et al (1998) is a good example 
 of what could be expected from redshifts of arc(let)s: 
 from their Keck observations, they show that the straight arc in A2390
is actually constituted of two lensed galaxies aligned along 
 the same direction, one at $z=0.931$ (reported earlier 
 by Pell\'o et al 1991) and the other one at $z=1.033$. These 
observations confirm the early conclusions from multicolor 
photometry (Smail et al 1993) as well as from theoretical 
considerations (Kassiola, Kovner \& Blandford 1992) that this
 straight arc should be composed of two galaxies. \\
About 50 redshifts of arc(let)s have been measured so far. 
However, from this sample it is difficult to infer valuable information 
 on the redshift distribution of
 $B>25$ galaxies or to constrain evolution models of galaxies 
 because it is biased in an unknown way. 
 Since most of these objects are very faint, only arclets showing bright
spots on HST images revealing  star-forming regions are generally 
selected. These features, which  increase the probability to detect
 emission lines,  optimize the chance to get reliable redshifts but
 generate a sample of  arclets where star forming galaxies are
preferentialy selected. Furthermore, due to the steep slope of galaxy counts 
 beyond $B=25$ the
magnification bias  favors observations of blue galaxies rather
than red ones.
So, even if the spectroscopy of arclets is
crucial for the lens modeling and eventually to obtain the spectral
energy distribution of high-redshift galaxies, the 
 spectroscopic sample of arc(let)s must be handled carefully 
 and need a  detailed analysis of the selection function prior to
 statistical studies. In the meantime, it is important to focus on 
 getting a few spectra of extremely high-redshift galaxies ($z>5$) that could be 
 observable thanks to high magnification.
\begin{figure}
\centerline{
\psdraft{figure=mellfig11.eps,width=7cm}}
\caption{\label{cl1358.ps}{\it The giant arc detected in Cl1358+6245 is
the most distant arc ever observed (z=4.92). The strong magnification permits 
 to detect some inner structures in a lensed galaxy at $z \approx 5$. Many bright spots 
 are visible (Franx et al 1997). The brightest spot on the left is also visible 
in the Soifer et al (1998) $J$-band observations and seems to be a dense core.
} 
}
\end{figure}
\subsubsection{Redshift distribution from lensing-inversion}
When it is possible to recover the lensing potential with  a good
accuracy, the lensing equation can be  inverted in order 
 to send the lensed image back to its source plane.  The 
 shape of the source can in principle be 
predicted for any redshift beyond the lens position. The basic 
principle  of the lensing-inversion  
approach was initially discussed by  Kochanek (1990) and refined latter 
 by Kneib et al (1994, 1996). If the  
  shape of the galaxies sent back beyond the lens plane 
 is parameterized by the  quantity $\tau=(a^2-b^2)/2ab\  e^{2i \theta}$, 
then it is 
 easy to show that in  the weak 
lensing regime, the complex quantity $\taug_I$ and its projection on a
$y$-axis, $\tau_y$, writes (Kneib et al 1994, 1996):
\begin{equation}
\taug_I=\taug+\taug_S \ ; \  \tau_{S_{y}}=\tau_{I_{y}}
\end{equation}
where the subscript $I$ and $S$ refers to the image and the source,
respectively.  Therefore, $\tau_y$ is an invariant.  The conditional   
probability of 
 a source to be at redshift $z$, given the shape and the position of the
image, and for a given mass model  is  
\begin{equation}
p(z \vert model) ={p(\tau_{S_{x}};\tau_{S_{y}}) \over p(\tau_{S_{y})}}
\ .
\end{equation}
For a simple distribution of the shape of the sources, it turns out
that this probability is maximum at the redshift where 
the deformation of the source is minimized. Therefore, the 
 lensing-inversion  predicts that the most probable redshift, for a 
given model,  
 is where the unlensed galaxies have a minimum 
ellipticity. This 
 intuitive assumption  proposed in Kneib et al (1994) was
 established on an observational basis by Kneib et al (1996) using the
HST-MDF galaxies as a fair sample of unlensed sources.
 The obvious interest of this
method is that it does not depend on the magnitude of the arclet but on
its position and its shape in the image plane. Potentially, it provides
 the redshift of any arclet up to the limiting magnitude of the observations.\\
The lensing inversion was first applied on A370  
(Kochanek 1990, Kneib et al 1994),  from the lens
modeling of the giant arc and some multiple images. Though the 
(unlensed) magnitude-redshift diagram found
for these arclets shows a good continuity with the faint spectroscopic
surveys (Mellier 1997), some of the predicted redshifts are  uncertain.
 In fact, as it is shown in Fort \& Mellier (1994, Figure 12), the X-ray 
isophotes and
the arclet positions do not follow the expectations of the lens
modeling of the eastern region. This is an indication that
while the modeling is excellent in the cluster center, the mass
distribution does not have a simple geometry beyond the giant arc and 
 therefore the lens model in this region is uncertain. 
 Similar complex substructures could exist on scales below 
 the resolution of the mass maps and could also produce  
 wrong redshift estimates.
Furthermore, the lensing inversion is also sensitive to the accuracy of
the shape measurements of each arclet, which in the case of very 
  faint objects could be an important source of uncertainty. \\
There are two solutions to solve these issues: first, it is highly
preferable to use HST images instead of ground based images. The 
 recent spectroscopic confirmations by Ebbels et al (1996, 1998) 
 of most of the redshifts predicted from the lensing-inversion in A2218 
 from the HST data (Kneib et al 1996) are wonderful
demonstrations of the capabilities of such a technique when used 
 with superb images.  Second, it is
important to focus on lensing-clusters with simple geometry in order to 
lower the uncertainties on the lens modeling. In this
respect, though A370 and A2218 are rather well modeled, they are not
the simplest, and clusters like MS0440, A1689 or
MS2137-23 appear as  better candidates.
\subsubsection{Probing source redshifts using various lens-planes}
A more natural and simple way to infer the redshift distribution of the
faint galaxies is to look for arc(let)s or weak lensing signal through
 a set of different lensing clusters having increasing redshifts. The 
ratio of lensed versus 
unlensed faint galaxies and the amplitude of the shear patterns 
 as function of redshift probe directly the spatial distribution 
 of the galaxies along the line of sight.  This idea was tentatively
explored by Smail et (1994) who analyzed the lensing signal in three
lensing clusters at redshifts 0.26, 0.55 and 0.89 respectively.  They 
found that most $I<25$ objects cannot be low-z dwarf galaxies and
concluded  that a large fraction of 
 $I=25$ galaxies are beyond $z=0.55$.  The absence
of any significant 
 lensing signal in the most distant cluster led to inconclusive
results on the fraction of these galaxies which could be at very large
redshift.  Fortunately, the distant clusters observed by  Luppino \& Kaiser
(1996) and Clowe et al (1998)  provided considerable insights 
 about the high-redshift tail of faint galaxies.  The detection of 
weak lensing in three $z>0.75$ clusters put strong constraints on
their samples of  $23.5<I<25.5$ galaxies, which must be dominated 
  by a $z>1$ population. These three lensing clusters strengthen 
 the conclusions   
obtained from the shear detected around Q2345+007 (Bonnet et al 1993)
 which is also produced by a high-redshift cluster (Mellier et al 1994, 
 Fischer et al 1994, Pell\'o et al 1996). \\
The use of distant clusters is 
 promising because it is a direct consequence
of the detection of weak lensing signal, regardless of the accuracy of
the mass reconstruction. The shape of the redshift distribution of the 
galaxies can be inferred if many clusters at different redshifts 
 map the lensing signal. Up to now, the number of clusters 
 is still low,  but it will continuously increase during the coming years. 
 It is however worth noting that the derived shape of the redshift 
distribution also depends on the accuracy of the lens modeling, as well
as on the dynamical evolution of clusters with look-back time. At high 
redshift it is possible that the lensing signal decreases rapidly, 
not only because of the absence of background sources, but also because
clusters of galaxies are no longer dense and massive to produce 
 gravitational distortion. It will be important to disentangle these 
two different processes.
\subsubsection{The distribution of faint galaxies from the
magnification bias }
When the slope of the galaxy number count is lower than $0.3$, a sharp 
decrease of the galaxy number density  is expected 
 close to the critical radius corresponding to the redshift
of the background sources (see Eq.(23)). For a broad redshift distribution, the
cumulative effect of each individual redshift 
  results in a shallow depletion area which 
spreads over two limiting radii 
 corresponding to  the smallest and the largest critical lines 
 of the  populations dominating the  redshift distribution.
   Therefore, the  shape and the 
 width of depletion curves reveal the redshift distribution of the 
 background sources, and their analysis should provide valuable 
 constrains on the distant galaxies.  Like the lensing inversion, this 
is a statistical method which also needs a very good 
 modeling of the lens; but in contrast to it, it does not need
 information on the shapes of arclets, so the
``depletion-redshift'' could be a more relevant approach for very faint
objects. \\
This method was first used by Fort et al (1997) in the cluster
Cl0024+1654 to study the
faint distant galaxy population in the extreme magnitude ranges 
$B=26.5-28$ and $I=25-26.5$. The (unlensed) galaxy number counts were first 
calibrated using CFHT blank fields and checked from comparison with 
 the HST-HDF counts.  In this magnitude range, the slopes are close to 0.2, so  
 these populations can produce a highly contrasted depletion area, so
they  are  best  suited for this project. In this cluster, 
the lower boundary of the depletion is sharp and the growth curve 
  toward the upper radius extends up to 60 arcseconds 
 from the cluster center, as expected if the high-redshift tail 
 is a significant fraction of the lensed galaxies. Fort et al  
 concluded that  $60\% \pm 10\%$ of the $B$-selected galaxies are  
between $z=0.9$ and $z=1.1$ while most of the remaining $40\% \pm 10\%$
galaxies appears to be broadly distributed around a redshift of
$z=3$. The
$I$ selected population shows a similar bi-modal distribution,
but spreads up to a larger redshift range  with about 20\% above  $z > 4$. \\
There is no yet spectroscopic confirmation that the double-shape redshift
distribution predicted by Fort et al is real. Indeed, there are still
 uncertainties related to this method: in the particular case of 
 Cl0024+1654, the redshift of the
 arc used  to scale the mass was assumed to be close to $1$.  We 
know from recent Keck spectroscopic observation that it is at redshift
1.66 (Broadhurst et al 1999), so the innermost
calibration of the sources has to be re-scaled. 
 The method is also sensitive to the lens modeling of the projected
mass density. In the case of Cl0024+1654, it is rather well constrained
from the subarcsecond ultra-deep images of Fort et al, and the 
predicted velocity dispersion is very close to the measured values
from the galaxy radial velocities (Dressler \& Gunn 1992).   \\
This approach was recently generalized by B\'ezecourt, Pell\'o \& Soucail  
(1998a),   
in order to predict the number counts  of arc(let)s 
 with a magnification larger than a lower limit.  The prediction need
first a model for galaxy evolution which reproduces some typical 
 features of field galaxies, like the observed galaxy counts and redshift 
distributions of spectroscopic surveys.  The best final model can then be 
used to predict the number of arc(let)s brighter than a lower 
  surface brightness limit and 
 with magnification larger than a lower limit, for any lensing cluster
 whose mass distribution can be modeled  properly. 
B\'ezecourt et al (1998a,b) used this technique 
 to build the best model capable to produce the number of arc(let)s 
 observed in A370 and A2218.  They predicted 
also a bimodal redshift distribution, but their expected number of giant
arcs is overestimated by a factor of two. This inconsistency 
 is still difficult to interpret. Indeed, the method should be handled 
 carefully, because it both depends on the modeling of the lens and on the
modeling of galaxy evolution. In particular, fitting of
 counts and redshift distributions of galaxies produces models 
 which have 
degeneracies (Charlot 1999) that are potential limitations.  Nevertheless, 
like the Fort et al method,  this
is an interesting and original idea which can certainly be improved in
the future using much better data.  As compared with the
lensing-inversion, it does not depend on the shapes of distant
galaxies, and just needs very deep counts (the 
 B\'ezecourt et al (1998a,b) method does need shapes of lensed galaxies, but they only use reasonably magnified
arc(let)s, so it is not a difficult, so it is not a difficulty). In the ``depletion-redshift'' technique this is a great 
advantage for the faintest (most distant?)
objects, because, as the HDF images show,  many of them have bright spots
but do not show regular morphologies,  so the lensing-inversion 
procedure could be inefficient for these galaxies.
 These very first attempts must be pursued on many lensing clusters in
order to
provide reliable results on the redshift distribution of the faintest
 galaxies. 
\begin{figure}
\centerline{
\psdraft{figure=mellfig12.eps,height=14cm,width=16cm}}
\caption{\label{distribz.ps} {\it Redshift distribution inferred from lensing
techniques. The top panel is the spectroscopic sample of arc(let)s compiled from 
the literature. The bottom panel includes the spectroscopic sample and the 
 redshift predictions of arclets from lensing inversion. On both histograms,  
the plots of the redshift distributions from the depletion curves in B (solid 
line) and 
I (dashed line) prediction by Fort et al (1997) are shown. The maximum of each 
curve is arbitrary fixed to the maximum of the histogram. }
}
\end{figure}
\vskip 5truemm
The redshift distribution obtained by these various techniques are 
summarized in Figure \ref{distribz.ps}.  The distribution is broad 
and the comparison with Figure 9 of Fort \& Mellier (1994) 
 shows that the median redshift and the width of the distribution 
  increases continuously.   The median redshift distribution of 
giant arcs was close to 0.4 four years ago and has  increased up  
 to 0.7 \ but with a more pronounced high-redshift tail.  
A significant fraction of the new redshifts exceeds 
 1.5, with a visible trends towards very high redshifts ($z>2.5$).  
The median redshift obtained from lensing inversions is close to 0.7, with a good 
correlation with other methods.  It is
surprising to see that the redshift distribution obtained by the
depletion curves predict two peaks which seem to be visible
in the redshift distribution of arc(let)s as well, both on the spectroscopic and 
 lensing inversion samples. Due to the 
somewhat different selections criteria used for these two samples, a 
resemblance was not really expected, and it may be an indication that selection 
 biases in the spectroscopic sample of arc(let)s are not critical.\\
Since the observations 
 of lensed galaxies use simultaneously  magnification, color 
selection, shape selection (elongation) and relative position 
 with respect to the cluster center (deviation angle), it is 
 thus possible to select jointly  drop-out galaxies with the radial-distance 
criterium 
 of elongated object in order to select extremely distant galaxies.
   Since this 
method seems to be efficient, it will be certainly applied to 
select samples of $z>5$ galaxies. This will probably be a main 
goal for the next years. 
\subsection{Spectral content of arc(let)s}
The strong magnification  of giant arcs also permits to study the 
content and star formation rates of high-redshift galaxies.  Preliminary
studies started with Mellier et al (1991) and 
 Smail et al (1993) who explored the spectral
content  of some arcs. These samples do not show
spectacular starburst galaxies and seem to be compatible with a 
continuous star
formation rate.  The HST images confirm that many of these galaxies
have bright spots with ongoing star formation. The star
formation rates inferred from new optical spectra of arcs in A2390
(B\'ezecourt \& Soucail 1996), in A2218 (Ebbels et al 1996), 
  in Cl1358+6245 (Franx et al 1997) 
 or in Cl0939 (Trager et al 1997) range from 5 
 to 20 $M_{\odot}$/yr and are consistent 
with other observations (Bechtold al 1998), but none of the rates 
 computed for arclets are corrected from dust extinction. \\
It is only very recently that the material of magnified arcs is 
being studied in detail. 
 Trager et al (1997) made the first attempt to estimate
the metallicity of the arclets at $z > 3$  detected in Cl0939 
 with the Keck telescope 
 and found that they are metal-poor systems, having $Z<0.1 Z_{\odot}$. 
 The very first CO observations at IRAM 
of the giant
arc in A370 (Casoli et al 1997: CO(J=2-1) detected)  and at Nobeyama 
Observatory in MS1512-cB58 (Nakanishi et al
1997: CO(J=3-2) undetected) have demonstrated that (sub)millimeter observations 
are feasible thanks to the magnification and can
provide useful diagnoses on the molecular and gas content of galaxies
at high redshifts.   If, as suggested by the Cosmic Infrared Background
(Puget et al 1996; see also Guiderdoni 1998 and references therein), 
 a significant fraction of the  UV emission of distant galaxies is released in the 
submillimeter range, the observations of lensed galaxies 
 in the submillimeter and millimeter bands could be a major step 
 in our understanding of the history of star formation in galaxies. 
  Blain (1997) emphasized that  the joint submillimeter 
 flux-density/redshift relation
and  the steep slope counts make the observations of
lensed distant galaxies in this waveband optimum, so a   
large number of bright (magnified) sources is expected.  Both SCUBA 
(at 450$\mu$m and  850$\mu$m)
and IRAM (at 1.3mm) can therefore benefit from magnification of distant lensed 
galaxies.  The large field of view and the wavelength range 
 of SCUBA at JCMT seem perfectly suited for this program. Smail, 
 Ivison \& Blain  (1997) are carrying out a long term program of 
observations of lensing clusters with this instrument. They detected sources 
in A370 and Cl2244-02 with a success rate which 
 implies that the number density of these galaxies is about 3 times higher
than expected from the 60 $\mu$m IRAS count. 
 Their observations of 
a new sample of 7 lensing clusters  (Smail et al 1998) show that
the energy emitted by these galaxies is much higher than the expectations from 
  nearby galaxies. Most of these
galaxies are at redshift larger than 1, and probably less than 
 5.5\ . The star formation activity of high-redshift galaxies 
 is therefore important and for those which have an optical
counterpart, their morphology reveals  
 signs of merging processes.  Therefore,  the  
  star formation activity seems frequently triggered by interactions 
 (Smail et al 1998), 
 which corroborates the recent ISOCAM observations in some giant arcs, like
 in A2390 (L\'emonon et al 1998). The star formation rates
 measured from the various fluxes have a very broad range,  
  between 50 to 1000 $M_{\odot}$/yr, but 
 they are  difficult to estimate accurately, in particular for the hyperluminous system
in A370 (Ivison et al 1998) because AGNs could contribute 
significantly to the flux.
 \\
The submillimeter observation is certainly one of the most promising 
tool for the next years.  The magnification 
and the  shape of the continuum produced by dust make 
 the ``submillimeter gravitational telescope''    
perfect to study the high-redshift Universe and the star formation
history of galaxies.  \\
\subsection{Morphology of highly-magnified galaxies}
Though the coupling of gravitational telescopes with the high resolution 
 images from  HST is superb to probe the intrinsic morphology of
 arc(let)s, no significant results raised from the last years studies.    
It turns out that many HST images of arcs are constituted of  
 two parallel elongated arcs, like in A2390 (Kneib et al in preparation) or 
MS2137-23 
(Hammer et al 1997; see Figure 2 of this review, the arcs A and C in
MS2137-23), which supports the idea that these distant galaxies
are interacting systems as it was reported in the previous section. 
 Others show bright spots which are interpreted as star forming regions,
 like Cl2244-02 (Hammer \& Rigaud 1989), A2390 (Pell\'o et al, in preparation) 
or 
Cl0024+1654 (Tyson et al). In the case of the giant arc in Cl1358+62 at $z=4.92$, 
 the comparion of the visible
and the near-infrared observations obtained with the HST and the Keck telescopes
  (Soifer et al 1998)  
reveals that one of its bright spots already contains half the stellar mass of the galaxy. 
 It shows that at this redshift galaxies may already have dense cores. Furthermore, 
the  visible and near infrared imaging and spectrospic data show also that reddening
produced by dust is important, even at that redshift.\\
The detailed description of a $z\approx5$-galaxy 
  done by Soifer et al  strengthens the usefulness of 
 image reconstruction techniques of high-redshift lensed galaxies, as the one initially 
proposed by Kochanek et al (1989). Some attempts have been done already in 
Cl0024+1654 (Wallington et al 1995; Colley, Tyson \& Turner 1996) or 
 in MS2137-23 (Hammer et al 1997). They succeed to reproduce a single image 
 in the source plane, but the details of the morphology are still
uncertain and do not provide yet valuable information on  
 distant galaxies. Therefore, it is still premature 
to present quantitative results on the distant galaxies, first because 
this sample is poor and incomplete, second because they are too much
uncertainties in the source reconstruction. 
\section{COSMOLOGICAL PARAMETERS}
The potential interest of gravitational lensing for cosmography,    
 already  discussed by Blandford \& Narayan (1992)   and
 Fort \& Mellier (1994) for the particular cases of 
arc(let)s, is clear but also challenging. The
increasing evidence that the whole set of observations are compatible 
with a non-zero cosmological constant for instance,  motivated 
 many new studies devoted to the 
constraints on $\lambda$,  
following the early suggestions by Paczynski \& Gorski (1981).  
 One of the most promising and reliable approaches explored by recent 
 theoretical studies has been raised by Bernardeau et al (1996) and 
 Van Waerbeke et al (1998). These works clearly demonstrate that 
 the determination of the cosmological parameters $(\Omega_0,\lambda)$ 
 using future surveys  of weak lensing induced by large-scale 
structures should provide $\Omega$ with a
  high accuracy (see Figure \ref{omegalambda1.ps} and Figure
\ref{omegalambda2.ps}).   Since it was already pointed out 
in section 4.1\ , I do not discuss it in more detail. In the 
following I focus on new and more speculative investigations, which are
still difficult to implement, but  seem promising in the future.\\
\subsection{Constraints from cluster reconstruction}
Following the early suggestion by Paczynski \& Gorski (1981) for
multiply imaged quasars,
 Breimer \& Sanders (1992) (see also Fort \& Mellier 1994, 
  Link \& Pierce 1998) emphasized that 
  the ratio of angular diameter distances of 
   arc(let)s having different  redshifts does not 
depend on the Hubble constant and therefore can constrain
 $(\Omega,\lambda)$. This ratio still depends on the 
 mass distribution  within the two critical lines corresponding to 
 redshifts $z_1$ and $z_2$,
 so it is worth noting that it is sensitive to the modeling of the lens. 
 It is only in the case of an isothermal sphere model that
  the radial positions of the critical lines $\theta$, where
arcs at a given redshift are formed, only depend on the angular
distances $D(z_s,\Omega_0,\lambda)$:
\begin{equation}
\left({\theta_1 \over \theta_2}\right)=\left({D_{LS}(z_1,\Omega_0,\lambda) 
 \over D_{OS}(z_1,\Omega_0,\lambda)}\right)\left({D_{OS}(z_2,\Omega_0,\lambda)
\over D_{LS}(z_2,\Omega_0,\lambda)}\right) \ .
\end{equation}
Since cluster potentials  are by far more
 complex than isothermal spheres, in practice the method 
 works only for very specific cases, like clusters with regular morphology,
  and if auxiliary independent data, like high-quality X-ray images or additional 
 multiple images,    
 help to constrain the lens model. So
far,  no case was  found where  the modeling of two (or more) arc 
systems at very different redshifts is sufficiently reliable. However,   
the joint HST images and spectroscopic redshifts obtained 
 with new giant telescopes
 should provide such   perfect configurations in a near future. A1689 
 or MS0440 seem
 good examples of such candidates because both show many arc(let)s 
 and have a regular shape. \\
A similar approach has been proposed by Hamana et al (1997),  using 
 the arc cB58 observed in the lensing cluster MS1512.4+3647.  Assuming
that the dark matter distribution is sufficiently constrained by the 
 ROSAT and ASCA data, the magnification and number of multiple 
images of cB58 only depend on the cosmology.  One should therefore 
use the detection of counter-image to cB58 to  constrain the domain 
 $(\Omega_0,\lambda)$ which cannot produce a counter-arc. This 
 point was discussed by Seitz et al (1997) who argue that in practice 
 it cannot work because it depends too much on the modeling of the
lensing cluster. The variation of the lensing strength as function of
cosmology is small, lower than 0.5\% between an EdS universe and
an $\Omega=0.3,\ \lambda=0$ universe.  Furthermore, the use of independent
X-ray data to model the dark matter demands a
very good understanding of the physics of the hot gas for each
individual cluster considered.  
\\ 
More recently, Lombardi \& Bertin (1998c) have proposed  
to use weak lensing inversion to recover simultaneously 
 the cluster mass distribution and the geometry of the Universe.   
The method assumes 
 that  the redshifts of the lensed
 galaxies are known.  In that case, for a given cosmology, it 
is possible to compute the shear at a given 
angular position which is produced on a  lensed galaxy located in 
a narrow redshift range, from the observed ellipticities of the 
galaxies at that angular position. Conversely, if the shear is known,
then it is possible to infer the best set of cosmological 
 parameters which reproduce the observed ellipticities of the galaxies.  
 Therefore, it   
is  possible to iterate a procedure, starting from an arbitrary 
guess for the set $(\Omega,\lambda)$, which at the final step 
will procure simultaneously 
the best mass inversion with the most probable $(\Omega,\lambda)$. The 
key point is the assumption that the redshift of each 
 individual source is known.  The method should
 provide significant results if at least a dozen of clusters with different 
 redshift are reconstructed using this iterative procedure 
 (Lombardi \& Bertin 1998c).  Indeed, this inversion is 
 demanding in telescope time since  a very good knowledge
of the redshifts of many lensed sources is necessary; but otherwise the method 
seems promising.
\subsection{Statistics of arc(let)s}
Up to now, several tens of multiply imaged distant galaxies  have
already
been detected in clusters of galaxies and this number will probably
increase by a large factor within the next years.   
 Since the fraction of rich clusters (and therefore
lensing-clusters) strongly depends on the 
  cosmological  scenario, we expect the number of arc(let)s  to
  depend on cosmological parameters (Wilson, Cole \& Frenk
1996b).  \\
It is well known that present-day statistical studies of clusters of 
 galaxies are limited by the few samples of cluster catalogs with 
 well understood selection function.  Wu \& Mao (1996) 
 analyzed the statistics of arcs in the EMSS cluster sample (Gioia et al 1990)
 and show that the fraction is twice the one expected for an EdS
Universe, but compatible with a flat $\lambda=0.7$ model (see also Cooray 1998). 
Unfortunately, the geometry of the
mass distribution and the substructures present in the lensing cluster
 increase the shear contribution to the magnification and  
 change dramatically the expected number of arcs 
 (Bartelmann, Steinmetz \& Weiss 1995; Bartelmann 1995b; 
 Hattori, Watanabe \& Yamashita 1997, Bartelmann et al 1998).
The importance
of accurate simulations of clusters is clearly obvious from the recent
studies.  Bartelmann et al (1998)  find a totally opposite result 
 to Wu \& Mao,
 and conclude that an open model ($\Omega=0.3$ and $\lambda=0$) is 
 preferred to any flat models to 
reproduce the number of arcs observed. The other models, including 
$\lambda=0.7$ model fail by about of factor of ten.  This 
 is a clear demonstration that the use of statistics of arc(let)s 
 to constrain cosmological scenario is very sensitive to 
 the assumptions. The constraint on $\lambda$ 
 from the fraction of arc(let)s  
 is therefore rather weak  and hopeless for the moment.  \\
\subsection{Magnification bias}
The depletion of the galaxy number density as function of radial 
distance from the cluster center can potentially provide 
information on the cosmological constant.  The reason for this 
is ultimately the same as for giant arcs, namely, the ratios of 
 angular distances which strongly depend on the cosmological constant.
 Therefore, if the redshift distribution of the sources and
 the mass distribution of the lensing cluster are known, the shape 
of the depletion curve, in particular, its extension at large radius 
 is constrained by $\lambda$. \\
 Fort et al (1997) have used 
 this property in order to constrain the cosmological constant. 
 They used ultra-deep images of the lensing clusters  Cl0024+1654 and
A370 which permit to have a good signal to noise ratio of the depletion
curves. These clusters have giant arcs with known redshift so the  
 mass at a given critical line can be scaled. 
 The method provide
jointly the redshift of the sources, and the cosmological parameters.  
 Fort et al (1997)  found that the location of this high redshift
critical line rather favors a flat cosmology with $\lambda$
   larger than 0.6. \\
 It is remarkable that from these two clusters only the method predicts   
 a value of $\lambda$ compatible with other independent approaches (see
 White 1998 and references therein).  
 Since it needs a good model for the lens, 
 this  method has still many uncertainties 
 and can be significantly improved with a large sample of  arc
clusters, in particular by using
a maximum  likelihood analysis applied to the probabilities of
reproducing
their observed local shears and convergences.
A strong improvement can come from the new
possibility to use the redshift distribution found independently. 
 This should be possible using photometric redshifts. 
\vskip 5truemm
All the methods described above do not yet provide convincing results on
$\lambda$ mainly because they use simultaneously different quantities 
 which are degenerate without external information: mass distribution
of the lensing-cluster, redshift distribution of the sources,
cosmological parameters, and evolution scenarios of clusters and
of sources. The approach using statistics of arc(let)s looks promising 
but demands very good simulations and a good understanding of selection
functions of cluster samples which are used for comparison with
observations.  The method using lens modeling needs very good lens models 
 and information on  the redshift distribution of galaxies, in particular for the most
distant ones, since they contain the population which depends the most on
$\lambda$.  This approach can use the redshift distribution obtained
 from photometric redshifts, and should focus on regular lensing clusters
containing giant arcs with known redshift. As emphasized by Fort et al 
(1997) and Lombardi \& Bertin (1998c), significant results cannot be 
expected until many clusters have been investigated. This should be done
within the next years, in particular using 10 meters class telescopes.  
However, it is remarkable that the Fort et al limit corresponds to the value 
 given by Im, Griffiths \& Ratnatunga (1997) from the measurement of 
 strong lensing produced by elliptical galaxies, and to the upper limit
given by Kochanek (1996) from the statistics of lensed quasars.

\section{LENSING the CMB}
The measurement of CMB fluctuations is a major goal 
for cosmology in the next decade (see White, Scott \& Silk 1994, and references
therein). The shape of the power spectrum over the whole spectral range
contains a huge amount of 
 information which permits to constrain the cosmological scenario
 with an incredibly high accuracy. However, the reliability of the
interpretation of the features visible on the spectrum needs a 
complete and detailed understanding of all the physical mechanisms 
responsible for its final shape.  Gravitational lensing induced by  
foreground systems along the line of sight may play a role, so 
it is important to predict in advance whether it can modify the 
 signal from the CMB and, if so, what are the expected amplitudes of 
 the effects.\\ 
 Since surface brightness is conserved by the gravitational lensing effect 
(Etherington 1933), only fluctuations of the CMB temperature maps 
can be affected by lensing. However, even for strong lenses, no
significant modification of the power spectrum is expected on large 
scales (Blanchard \& Schneider 1987) and therefore there is no hope to 
detect positive signals of the coupling between CMB and gravitational 
lensing  in the Cosmological Background Explorer (COBE) maps.  Nevertheless, 
 since  the COBE-DMR experiment has demonstrated that fluctuations 
 exist (Smooth et al 1992), the study of the lensing effect on smaller scales 
than COBE resolution is potentially interesting and has some advantages with 
respect to weak lensing on distant galaxies. First, contrary to lensed
galaxies, the redshift of the source, namely the last scattering surface,
 is well known and spreads over a very small redshift range. Second, 
with the on-going and the coming of high-resolution
 ground-based and balloon observations as well as the 
two survey satellites MAP and Planck-Surveyor,
observation of low-amplitude
 temperature distortions on small scales will become
  possible and will permit to investigate possible lensing effects.\\
Early theoretical expectations from Blanchard \& Schneider (1987) or
 Cole \& Efstathiou (1989) show that the {\it shape} of the small-scale 
temperature fluctuations can be modified by lensing effects; in 
particular they can redistribute the power in the power spectrum. 
 In contrast, the {\it amplitude }  
 of the temperature anisotropy on medium and small scales has been a matter of
debates during the last decade (see the review by Blandford \& Narayan
1992,
and more recently  Fukushige \& Makino 1994; 
Fukushige, Makino \& Ebisuzaki
1994, Cay\'on, Mart\'{\i}nez-Gonz\'alez \& Sanz 1993a,b, 1994). The
conclusions of these works showed strong discrepancies, depending on the
assumptions used to explore the deflection of photons and 
 to model inhomogeous universes. Furthermore, the 
 expectation values also depend on the
cosmological models. Indeed, the most recent critical studies 
  show 
that the effect of lensing on large scales is negligible 
 (Seljak 1996, Mart\'{\i}nez-Gonz\'alez, Sanz \& Cay\'on 1997).  In 
 particular, the non-linear evolution of the power spectrum does not 
 increases significantly the amplitude on these scales. On the other 
 hand,  the   
gravitational lensing effect reduces the power spectrum on small scales, 
and eventually can smooth out acoustic peaks on scales below $l\approx
2000$ (Seljak 1996).   Mart\'{\i}nez-Gonz\'alez et al (1994) obtained similar 
conclusions as Seljak,  
  that is, the contribution of lensing is small but not negligible and 
should be taken into account in the detailed analysis of future 
CMB maps. Furthermore, the transfer of power from large to
 small scales induces an important increase of power 
 in the damping tail, which results in  a decrease of very small 
 scale amplitudes   at a smaller rate than expected without lensing
 (Metcalf \& Silk 1997, 1998).
 According to Metcalf \& Silk (1997), 30\% of the 
 additional power at $l=3000$ comes from $l<1000$ scales, and 8\% 
 from $l<500$ in the case of a $\sigma_8=0.6$ model.\\
Zaldarriaga \& Seljak (1998) pointed out that 
gravitational lensing does not only smooth the temperature anisotropy,
but can also change the polarization.  The
polarization spectra are more sensitive to gravitational lensing effects  
 than  the power spectrum of the temperature because the acoustic
oscillations of polarization spectra have sharper oscillations
 and  can be smoothed out more efficiently by lensing 
 than temperature fluctuations. The effect is 
 small but can reach amplitudes of about 10\% for $l<1000$ scales. More 
remarkably, because of the coupling between $E-$type and $B-$type polarizations (Seljak
1997b),
 gravitational lensing can generate low amplitude $B$-type polarization, even
if none is predicted from primary fluctuations (for instance, for scalar perturbations). \\
Because the signal is weak and only concerns the small scales, 
 temperature and polarization fluctuations induced by gravitational
lensing will be difficult to measure with high accuracy and seems 
 a hopeless task before the Planck-Surveyor mission.  It is therefore 
 valuable to explore alternatives which could provide better or 
complementary information which couples lensing and CMB.  
An interesting idea is to analyze the  non-Gaussian 
 features induced by the displacement fields generated  by 
gravitational lensing on the CMB maps (Bernardeau 1997, 1998b).  
 As for weak lensing on distant galaxies,            
 the CMB temperature map can be sheared and magnified.
The resulting  
 distortion patterns are direct signatures of the coupling between the CMB 
and the foreground lenses. Bernardeau (1998b)
argued that the distortion map produced by lensing can be decoupled 
 from other fluctuation patterns because it generates similar
magnification and deformation on close temperature patches, 
 which therefore can be correlated.  
 He also explored the consequences of the non-Gaussian 
signal on the four-point correlation function. Unfortunately, the 
signal is very small, and it is even not clear on which scale the
signal is  highest, in particular because the non-linear evolution of
the power spectrum was not considered by Bernardeau.  The weakness of 
 the signal, and the fact that the four-point 
correlation function could be contaminated by other non-Gaussian 
features are strong limitations of Bernardeau's suggestion.
  Therefore, Bernardeau (1998b) preferred to focus on  
 the modification of the ellipticity distribution function of the
temperature patches  induced by lensing. From his simulated lensed maps,
a clear change of the topology of the temperature maps is visible: 
the shapes of the structures are modified and their contours look  sharper   
 than for the unlensed maps. However, the signal is still marginally
detectable on a 10$^o\times$10$^o$ map, even with Planck-Surveyor. \\
From these investigations it is clear that weak lensing on the CMB  has 
small effects on the spectrum
of the temperature and polarization power spectra and on the 
non-Gaussianity of the CMB temperature maps. However, with typical 
amplitude of 1\% to 10\% percent they can be detected with future 
missions, so they must be taken into account for detailed investigations
of the CMB anisotropy on small scales.  This is an important prediction
  since the detection of gravitational
lensing perturbations of the CMB will be possible with Plank-Surveyor.
 Its high sensitivity and spatial resolution are 
 sufficient  to permit to break the geometrical degeneracy 
expected from linear theory, and to disentangle different
($\Omega,\lambda$) common models (Metcalf \& Silk 1998; Stompor 
\& Efstathiou 1998).  It is worth noting that 
 these  analyses can be used jointly with the 
 weak lensing  maps of large scales structures on background galaxies 
 which will also provide ($\Omega,\lambda$) with a very good accuracy.
\section{FUTURE PROSPECTS}
Though the use of weak lensing analysis and its applications in
cosmology made spectacular progress during the last five years, 
 most of the astrophysical questions 
addressed by Fort \& Mellier (1994) in their conclusions are still pending.    
However, it seems that we are progressing quickly in the right direction, even 
 if some of these problems  are complex and should be envisioned  in a long-term perspective.\\
  The HST  images have dominated
most of the results, in particular in the modeling of clusters of
galaxies. Thanks to the formidable work devoted to mass reconstruction,  
 the projected mass density of clusters of galaxies are now robust and 
 reliable. It is now important  to couple strong and weak lensing 
features 
 (Seitz et al 1998, AbdelSalam 1998b, Dye \& Taylor 1998, Van Kampen 1998) in order 
 to build consistent models for clusters.
   It is worth noting that 
 for many of the issues discussed in this review, it was emphasised 
that precise and reliable mass reconstructions of clusters 
of galaxies are crucial and determine the reliability of many  
scientific outcomes.  In this respect, it is important to keep in mind that the 
redshift distribution of the sources is indispensable and that the new giant telescopes
will be the best tools for this purpose. \\
From the sample of clusters already analysed, there are converging results that
 $\Omega<0.3$ with a high confidence level. However, complete cluster samples are 
necessary for deeper investigations of cluster properties. They should come out 
rapidly from weak lensing studies of ROSAT samples (Rosati 1999).  
Indeed, we have now a very good understanding of the mass
distribution of each component (dark matter, hot gas, galaxies) in clusters of galaxies 
and we are close to understand    
the discrepancy between the lensing mass and the X-ray mass of clusters.  
During the next five years one can reasonably expect significant  improvement on our
knowledge of the dynamics of clusters of galaxies by using jointly weak lensing 
reconstruction, from HST and giant telescopes images, and 
 a full description of the hot gas, from AXAF and XMM observations. \\
In contrast, the investigation of galaxy halos from galaxy-galaxy lensing 
is still at 
its infancy and the preliminary results presented in this review must be 
 confirmed.  A  new generation of instruments will contribute to the 
development of this hot topic.   Below  
 10 arcseconds down to 2 arcseconds,  ``wide field'' HST observations
with the new Advanced Camera  devoted to deep galaxy-galaxy lensing
 studies appear to be a promizing approach. Beyond this scale, the high image 
quality of telescopes like 
 Keck, GEMINI, the VLT or CFHT will be  decisive to obtain valuable constraints from
 galaxy-galaxy lensing analysis between 10 to 60 arcseconds.  
\\
In parallel to these studies, we can now envision to fully exploit one 
of the most valuable information that gravitational lensing can provide, namely
 the relation(s) between light and mass distributions in the Universe.   
Theoretical studies have demonstrated that in a near future the weak lensing 
analysis coupled with the study of the galaxy distribution will allow to 
 understand  the evolution of the biasing factor with scale and redshift. However, 
it is important to  explore the case of non-linear and stochastic 
 biasing in order to  
 understand what parameters  can be reasonably constrained. 
  The possible existence of big dark halos around galaxies or in clusters   
of galaxies is also an unknow but fascinating topic. In this respect,
 the dark cluster candidate discussed by Hattori et al (1997), or
 the remarkable distortion field detected by Bonnet et al (1994) in the periphery of
Cl0024+1654, which does not seem to be correlated to luminous matter,
 should deserve more careful investigations.\\
The study of the contents and the past history of galaxies made 
 formidable progress as well.  It is clear that using jointly magnification 
of cluster-lenses with 
 the unprecedented image quality of HST, or with the wide field and high 
sensitivity of SCUBA are highly competitive tools. 
  In the future, continuous developments are 
expected, but the observation of extremely high redshift galaxies which could 
not
 be observed without magnification is certainly a major goal, in particular in 
the 
submillimeter wavebands.   As demonstrated by the recent study of Soifer
et al (1998), the coming of optical and near-infrared 
 spectroscopic capabilities on the giant telescopes will permit to 
 study in detail their spectral energy distribution and the kinematics
of their stellar and gas components.   
 The magnification permits to see a huge amount of details on the 
images of these 
 galaxies and we can envision to probe small details of these galaxies
from image reconstruction ``\`a la Kochanek''. Unfortunately, 
  though theoretical tools have been developed in order to 
 recover the morphology of these lensed galaxies, the quality of optical and
submillimeter data are not good enough to produce reliable details of the sources   
from inversion.  This is probably a goal for the Large Southern Array (LSA) which 
will have much better 
sensitivity and resolution.\\
With new instruments, like MEGACAM at CFHT (Boulade et al 1998) or the 
VST at the European Southern Observatory (Arnaboldi
et al 1999) in Paranal,
 we are now entering  the era of wide field subarcsecond imaging surveys 
 which will produce the first shear-limited samples or, similarly, the first 
 mass selected catalogues of gravitational condensations. Their design are optimized to
 investigate weak lensing induced by large scale structures in order to  
 produce the first mass maps of the Universe. They will permit 
 to recover the detailed spectrum of the projected power spectrum of mass density
fluctuations as well as to measure   $(\Omega,\lambda)$ with an accuracy
better than 10\% (see Figure \ref{omegalambda2.ps}). There are  still some theoretical 
issues (see sections
3.3 and 4.4)
which must be addressed in detail from both the theoretical and
simulation points of views.  For most of them, there are no crucial 
 conceptual difficulties, so they should be fixed rapidly. On the other
hand, the control of systematics which can affect weak lensing
measurement, as well as the correction for an anisotropic PSF could be 
 critical and should be considered seriously in the future for very low shear amplitudes   
($<1$\%). Nevertheless, these cameras, as well as the very wide field surveys 
 of the VLA-FIRST and the SDSS, should provide a major breakthrough in 
weak lensing applications for cosmology. \\
\begin{figure}
\centerline{
\psdraft{figure=mellfig13.eps,width=7cm}}
\caption{\label{omegalambda2.ps}{\it Constraints on $(\Omega,\lambda)$ from 
a weak lensing survey covering 10$\times$10 square-degrees. The bright
and dark regions refer to $1\sigma$ and $2\sigma$ level. The left strip   
is for an $\Omega=0.3$ universe, and the  right band for and $\Omega=1$
universe. The solid and the dot-dashed lines correspond to a
zero-curvature universe and  to a fixed deceleration parameter,   
respectively (from Van Waerbeke et al 1998).}
}
\end{figure}
On the longer term, after the crucial results expected for mass maps   
with wide-field CCD cameras, the New Generation Space Telescope (NGST) 
and Planck Surveyor could be  
 ultimate steps of this area.   The potential interest of NGST
 for weak lensing has been summarized by Schneider \& Kneib
(1998) who argued that low-mass clusters and groups 
of galaxies, as well as very distant clusters 
 should be detectable with this telescope. In parallel,  
 as reported by Stompor \& Efstathiou 
 (1998) and Metcalf \& Silk (1998), weak lensing on the CMB should be 
able to break the geometrical degeneracy and therefore to provide 
 ultimately the ($\Omega,\lambda$) parameters.  This coupling
between observations of CMB fluctuations and weak lensing analyses 
 emerges  as a  consecration illustrating the major roles played
by  these two complementary approaches to present-day cosmology.

\section*{{\small ACKNOWLEDGMENTS}}
I am particularly grateful to B. Fort for his advices, and  his friendly 
and continuous encouragements during the long periode of preparation 
 of the review. 
I would like to thank first F. Bernardeau, F. Casoli, S. Charlot, R. Ellis, 
  P. Schneider  and L. Van Waerbeke
 for their careful reading and useful comments  of the manuscript as well as 
for their strong support during its writing. 
I thank all the other close collaborators who participate to our   
 gravitational lensing projects, and the colleagues with who we had many fruitful 
 discussions,  
namely M. Bartelmann, H. Bonnet, T. Broadhurst, J.-C. Cuillandre, T. Erben, 
H. Hoekstra, B. Jain, N. Kaiser, J.-P. Kneib, C. Kochanek, J.-F. Le Borgne, P.-Y. Longaretti, 
G.  Luppino, R. Pell\'o, M. Pierre, C. Seitz,  S. Seitz, U. Seljak,   
  I. Smail, G. Soucail and G. Squires.  I thank I. Gioia for providing
me data prior to publication, and T. Brainerd, B. Jain, P. 
Schneider and G. Squires, for giving 
me their authorization to publish figures of their papers in this review. I thank 
especially, M. Dantel-Fort for her assistance and also for the crucial   
 work she does in order to have all these data set ready for our scientific objectives.
Part of this work was supported by the Programme National de Cosmologie which is funded by the
Centre National de la Recherche Scientifique, the Commissariat \`a l'\'Energie Atomique
 and the Centre National d'\'Etudes Spatiales, under the responsibility of the Institut 
National des Sciences de l'Univers  and the Indo-French Centre for the
Promotion of
Advanced Research IFCPAR grant 1410-2..
\newpage
\noindent {\small {\it Literature Cited}}
\vskip 3truemm
\small{
\noindent AbdelSalam, H. M., Saha, P., Williams, L. L. R. 1998a. {\it MNRAS}
 294:734-46.\\
\noindent AbdelSalam, H. M., Saha, P., Williams, L. L. R. 1998b. 
{\it Astron. J.} 116:1541-52.\\
\noindent Allen, S. W. 1998. {\it MNRAS} 296:392-406.\\
\noindent Amram, P., Sullivan, W.T., Balkowski, C., Marcelin, M., Cayatte, V. 1993.
 {\it Ap. J. Lett.} 403:L59-L62.\\
\noindent Arnaboldi, M., Capaccioli, M., Mancini, D., Rafanelli, P., 
Sedmak, G., Scaramella, R., Vettolani, G. P. 1999. In {\it Wide Field
Surveys in Cosmology}, ed. S. Colombi, Y.  Mellier, B. Raban. Fronti\`eres. \\
\noindent Babul, A., Lee, M. H. 1991. {\it MNRAS} 250:407-13.\\
\noindent Bar-Kana, R. 1996. {\it Ap. J.} 486:17-27.\\
\noindent Bartelmann, M. 1995a. {\it Astron. Astrophys.} 298:661-71.\\
\noindent Bartelmann, M. 1995b. {\it Astron. Astrophys.} 299:11-16.\\
\noindent Bartelmann, M. 1995c. {\it Astron. Astrophys.} 303:643-55.\\
\noindent Bartelmann, M. 1996. {\it Astron. Astrophys.} 313:697-702.\\
\noindent Bartelmann, M., Narayan, R. 1995. {\it Ap. J.} 451:60-75.\\
\noindent Bartelmann, M., Narayan, R., Seitz, S., Schneider, P. 1996.
{\it Ap. J. Lett.} 464:L115-L118.\\
\noindent Bartelmann, M., Schneider, P. 1993a. {\it Astron. Astrophys.} 268:1-13.\\
\noindent Bartelmann, M., Schneider, P. 1993b, {\it Astron. Astrophys.} 271:421-24.\\
\noindent Bartelmann, M., Schneider, P. 1994, {\it Astron. Astrophys.} 284:1-11.\\
\noindent Bartelmann, M., Schneider, P. Hasinger, G. 1994, {\it Astron. Astrophys.} 290:399-411.\\
\noindent Bartelmann, M., Steinmetz, M., Weiss, J. 1995. 
 {\it Astron. Astrophys.} 297:1-12.\\
\noindent Bartelmann, M., Huss, A., Colberg, J. M., Jenkins, A. Pearce,
F. R. 1998. {\it Astron. Astrophys.} 330:1-9.\\
\noindent Bechtold, J., Elston, R., Yee., H.K.C., Ellingson, E., Cutri,
R.M. 1998, in {\it The Young Universe}, ed. S.
D'Odorico, A. Fontana, E. Giallongo, pp. 241-48. PASP Conf. Series. Vol. 416.
\\ 
\noindent Ben\'{\i}tez, N., Mart\'{\i}nez-Gonz\'alez, E. 1997. {\it Ap. J.} 477:27-35.\\
\noindent Bernardeau, F. 1997a. {\it Astron. Astrophys.} 324:15-26.\\
\noindent Bernardeau, F. 1998a. {\it Astron. Astrophys.} 338:375-82.\\
\noindent Bernardeau, F. 1998b. {\it Astron. Astrophys.} 338:767-76.\\
\noindent Bernardeau, F., Van Waerbeke, L., Mellier, Y. 1997. {\it Astron. Astrophys.}
 322:1-18.\\
\noindent B\'ezecourt, J., Soucail, G. 1997. 
{\it Astron. Astrophys.} 317:661-69.\\
\noindent B\'ezecourt, J., Pell\'o, R., Soucail, G. 1998. {\it Astron. Astrophys.}
 330:399-411.\\
\noindent B\'ezecourt, J., Kneib, J.-P., Soucail, G., Ebbels, T.M.D.
1998b. Preprint astro-ph/9810109.\\
\noindent Blain, A. 1997. {\it MNRAS} 290:553-65.\\
\noindent Blanchard, A., Schneider, J. 1987. {\it Astron. Astrophys.} 184:1-6.\\
\noindent Blandford, R.D. 1990. {\it Q. Jl. R. Astr. Soc.} 31:305-31.\\
\noindent Blandford, R.D., Jaroszy\'nski, M. 1981. {\it Ap. J.} 246:1-12.\\
\noindent Blandford, R.D., Saust, A.B., Brainerd, T.G., Villumsen, J.V.
 1991. {\it MNRAS} 251:600-27.\\
\noindent Blandford, R.D., Narayan, R. 1992. {\it Annu. Rev. Astron. Astrophys.} 30:311-58.\\
\noindent Bonnet, H., Fort, B., Kneib, J.-P., Mellier, Y., Soucail, G.
1993. {\it Astron. Astrophys. Lett.} 280:L7-L10.\\
\noindent Bonnet, H., Mellier, Y., Fort, B. 1994. {\it Ap. J. Lett.} 427:L83-L86.\\
\noindent Bonnet, H., Mellier, Y. 1995. {\it Astron. Astrophys.} 303:331-44.\\
\noindent B\"ohringer, H., Tanaka, Y., Mushotzky, R. F., Ikebe, Y.,
Hattori, M. 1998. {\it Astron. Astrophys.} 334:789-98.\\
\noindent Boulade, O., Vigroux, L., Charlot, X., de Kat, J., Borgeaud,
P., Rouss\'e, J.Y., Mellier, Y., Gigan, P., Crampton, D. 1998. 
{\it Astronomical Telescopes and Instrumentation}. SPIE Vol. 3355.\\
\noindent Bouchet, F. 1996. Preprint astro-ph/9603013.\\
\noindent Bower, R. G., Smail I. 1997. {\it MNRAS} 290:292-302.\\
\noindent Brainerd, T. G., Blandford, R. D., Smail, I. 1996. 
{\it Ap. J.}  466:623-37.\\
\noindent Breimer, T. G., Sanders, R. H. 1992. {\it MNRAS} 257:97-104.\\
\noindent Bridle, S. L., Hobson, M. P., Lasenby, A. N., Saunders, R.
1998. {\it MNRAS} 299:895-903.\\
\noindent Broadhurst, T. 1995. Preprint astro-ph/9511150.\\ 
\noindent Broadhurst, T., Taylor, A. N., Peacock, J. 1995. 
{\it Ap. J.} 438:49-61.\\
\noindent Broadhurst, T., Huang, X., Frye, B., Ellis, R.S., Morris, J. 1999. Preprint.\\
\noindent Casoli, F., Encrenaz, P., Fort, B., Boiss\'e, P., Mellier, Y.
1996. {\it Astron. Astrophys. Lett.} 306:L41-L44.\\
\noindent Cay\'on, L., Mart\'{\i}nez-Gonz\'alez, E., Sanz, J. L.  1993a.
 {\it Ap. J.} 403:471-75.\\
\noindent Cay\'on, L., Mart\'{\i}nez-Gonz\'alez, E., Sanz, J. L.  1993b.
 {\it Ap. J.} 413:10-13.\\
\noindent Cay\'on, L., Mart\'{\i}nez-Gonz\'alez, E., Sanz, J. L.  1994.
 {\it Astron. Astrophys.} 284:719-23.\\
\noindent Charlot, S. 1999. In {\it The Next Generation Space Telescope: Science Drivers and
Technological Challenges}, eds. P. Benvenuti et al. ESA SP-429.\\
\noindent Clowe, D., Luppino, G. A., Kaiser, N., Henry, J. P., Gioia, I.
M. 1998. {\it Ap. J. Lett.} 497:L61-L64.\\
\noindent Cole, S., Efstathiou, G. 1989. {\it MNRAS} 239:195-200.\\
\noindent Colombi, S., Szapudi, I., Szalay, A. 1998. {\it MNRAS} 296:253-74.\\
\noindent Colley, W.N., Tyson, J.A., Turner, E. 1996. {\it Ap. J. Lett.}
461:L81-L86.\\
\noindent Connolly, A.J., Csabai, I., Szalay, A., Koo, D.C., Kron, R.
G., Munn, J. A. 1995. {\it Astron. J.} 110:2655-64.\\
\noindent Cooray, A.R. 1998. Preprint astro-ph/9801148.\\
\noindent Coutts, A. 1996. {\it MNRAS} 278:87-94.\\
\noindent Cuillandre, J.-C., Mellier, Y., Dupin, J.-P., Tilloles, P., Murowinski, R.,
 Crampton, D., Woof, R., Luppino, G.A. 196. {\it Publ. Astron. Soc. Japan}
108:1120-28.\\
\noindent Dekel, A., Lahav, O. 1998. Preprint astro-ph/9806193.\\
\noindent Dell'Antonio, I. P., Tyson, J. A. 1996. {\it Ap. J. Lett.} 473:L17-L20.\\
\noindent Deltorn, J.-M., Le F\`evre, O., Crampton, D., Dickinson, M.
1997. {\it Ap. J. Lett.} 483:L21-L24.\\
\noindent Dolag, K., Bartelmann, M. 1997. {\it MNRAS} 291:446-54.\\
\noindent Dressler, A., Gunn, J. E. 1992. {\it Ap. J. Supp.} 78:1-60.\\
\noindent Dressler, A., Oemler, A., Sparks, W. B., Lucas, R. A. 1994. 
 {\it Ap. J. Lett.} 435:L23-L26.\\
\noindent Donahue, M., Voit, M., Gioia, I.M., Hughes, J.,  Stocke, J.
199. {\it  Ap. J.} 502: 550-57.\\
\noindent Dye, S., Taylor, A. 1998. {\it MNRAS} 300:L23-L28.\\
\noindent Ebbels, T., Le Borgne, J.-F., Pell\'o, R., Ellis, R. S., 
Kneib, J.-P., Smail, I., Sanahuja, B. 1996. {\it MNRAS} 281:75-81.\\
\noindent Ebbels, T., Le Borgne, J.-F., Pell\'o, R., Ellis, R. S., 
Kneib, J.-P., Smail, I., Sanahuja, B. 1998. {\it MNRAS} 295:75-91.\\
\noindent Ellis, R.S. 1997. {\it Annu. Rev. Astron.
Astrophys.} 35:389-444.\\
\noindent Etherington, I. M. M. 1933. {\it Phil. Mag. } 15:761-75.\\
\noindent Evans, N. W., Wilkinson, M. I. 1998. {\it MNRAS} 296:800-12.\\
\noindent Fahlman, G., Kaiser, N., Squires, G., Woods, D. 1994. {\it Ap. J.} 437:56-62.\\
\noindent  Fischer, P., Tyson, J. A. 1997. {\it Astron. J.} 114:14-24.\\
\noindent  Fischer, P., Tyson, J. A., Bernstein, G. M., Guhathakurta, P.
1994. {\it Ap. J. Lett.} 431:L71-L74.\\
\noindent  Fischer, P., Bernstein, G., Rhee, G., Tyson, J. A. 1997. {\it Astron. J.}
 113:521-30.\\
\noindent  Fort, B., Prieur, J.-L., Mathez, G., Mellier, Y., Soucail, G.
1988. {\it Astron. Astrophys. Lett.} 200:L17-L20.\\
\noindent Fort, B., Mellier, Y. 1994. {\it Astron. Astrophys. Rev.} 5:239-92.\\
\noindent Fort, B., Mellier, Y., Dantel-Fort, M., Bonnet, H., Kneib,
J.-P. 1996. {\it Astron. Astrophys.} 310:705-14.\\
\noindent Fort, B., Mellier, Y., Dantel-Fort 1997. {\it Astron. Astrophys.} 321:353-62.\\
\noindent Franx, M., Illingworth, G. D., Kelson, D. D., Van Dokkum, P.
G., Tran, K.-V. 1997. {\it Ap. J. Lett.} 486:L75-L78.\\
\noindent Frye, B. L., Broadhurst, T. J. 1998. {\it Ap. J. Lett.} 499:L115-L118.\\
\noindent Frye, B. L., Broadhurst, T. J., Spinrad, H., Bunker, A. 1998.
Preprint astro-ph/9803066.\\
\noindent Fugmann W. 1990. {\it Astron. Astrophys.} 240:11-21.\\
\noindent Fukushige, T., Makino, J. 1994. {\it Ap. J. Lett.} 436:L111-L114.\\
\noindent Fukushige, T., Makino, J., Ebisuzaki, T. 1994. {\it Ap. J. Lett.}
436:L107-L110.\\
\noindent Garrido, J.L., Battaner, E., S\'anchez-Saavedra, M.L.,
Florido, E. 1993. {\it Astron. Astrophys.} 271:84-88.\\
\noindent Geiger, B., Schneider, P. 1997. Preprint astro-ph/9805034.\\
\noindent Geiger, B., Schneider, P. 1998. {\it MNRAS} 295:497-510.\\
\noindent Gioia, I. M., Shaya, E. J., Le F\`evre, O., Falco, E. E.,
Luppino, G., Hammer, F. 1998. {\it Ap. J.} 497:573-86.\\
\noindent Gioia, I.M., Maccacaro, T., Schild, R.E., Wolter, A., Stocke, J.T., Morris,
S.L., Henry, J.P. 1990. {\it Ap, J. Suppl.} 72:567-619.\\
\noindent Gioia, I.M.,Henry, J.P., Mullis, C.R., Ebeling, H. and Wolter,
A. 1998. submitted to {\it Astron. J}. \\
\noindent Girardi, M., Fadda, D., Escalera, E., Giuricin, G.,
Mardirossian, F., Mezzetti, M. 1997. {\it Ap. J.} 490:56-62.\\
\noindent Gorenstein, M.V., Falco, E.E., Shapiro, I.I. 1988. {\it Ap. J.} 327:693-711.\\
\noindent Griffiths, R. E., Casertano, S., Im, M., Ratnatunga, K.
U. 1996. {\it MNRAS} 282:1159-64.\\
\noindent Im, M., Griffiths, R. E., Ratnatunga, K. U. 1997. {\it Ap. J.} 475:457-61.\\
\noindent Guiderdoni, B. 1998, in {\it The Young Universe}, ed S.
D'Odorico, A. Fontana, E. Giallongo., pp. 283-88. PASP Conf. Series. Vol. 416.
\\ 
\noindent Gunn, J. E. 1967. {\it Ap. J.} 150:737-53.\\
\noindent Hamana, T., Hattori, M., Ebeling, H., Henry, P., Futumase, T.,
Shioya 1997. {\it Ap. J.} 484:574-580.\\
\noindent Hammer, F., Rigaud, F. 1989. {\it Astron. Astrophys.} 226:45-56.\\
\noindent Hammer, F., Teyssandier, P., Shaya, E. J., Gioia, I. M.,
Luppino, G. A. 1997. {\it Ap. J.} 491:477-82.\\
\noindent Hattori, M., Ibeke, Y., Asaoka, I., Takeshima, T., B\"ohringer, H., 
 Mihara, T., Neumann, D.M., Schindler, S., Tsuru, T., Tamura, T. 1997. 
{\it Nature} 388:146-48.\\
\noindent Hattori, M., Watanabe, K., Yamashita, K. 1997. 
 {\it Astron. Astrophys.} 319:764-80.\\
\noindent Hoekstra, H., Franx, M., Kuijken, K., Squires, G. 1998.
{\it Ap. J} 504:636-60.\\
\noindent Hudson, M. J., Gwyn, S., Dahle, H., Kaiser, N. 1998. 
{\it Ap. J} 503:531-42.\\
\noindent Ivison, R. J., Smail, I., Le Borgne, J.-F., Blain, A. W.,
Kneib, J.-P., B\'ezecourt, J., Kerr, T. H., Davies, J. K. 1998.
{\it MNRAS} 298:583-93.\\
\noindent Jain, B., Seljak, U. 1997. {\it Ap. J.} 484:560-73.\\
\noindent Jain, B., Seljak, U., White, S. 1998. Preprint
astro-ph/9804238.\\
\noindent Kaiser, N. 1992. {\it Ap. J.} 388:272-86.\\
\noindent Kaiser, N. 1995. {\it Ap. J. Lett.} 439:L1-L3.\\
\noindent Kaiser, N. 1998. {\it Ap. J.} 498:26-42.\\
\noindent Kaiser, N., Squires, G. 1993. {\it Ap. J.} 404:441-50.\\
\noindent Kaiser, N., Squires, G., Broadhurst, T. 1995. {\it Ap. J.} 449:460-75.\\
\noindent Kaiser, N., Luppino, G., Kofman, L., Gioia, I., Metzger,
Dahle, H. 1998.  Preprint astro-ph/9809268.\\
\noindent Kamionkowski, M., Babul, A., Cress, C. M., R\'efr\'egier, A.
1997. Preprint astro-ph/9712030.\\
\noindent Kasiola, A., Kovner, I., Fort, B. 1992. {\it Ap. J.} 400:41-57.\\
\noindent Kasiola, A., Kovner, I., Blandford, R. D. 1992. {\it Ap. J.} 396:10-19.\\
\noindent Kneib, J.-P., Mellier, Y., Fort, B., Mathez, G. 1993. 
{\it Astron. Astrophys.}  273:367-76.\\
\noindent Kneib, J.-P., Mathez, G. Fort, B., Mellier, Y., Soucail, 
G., Longaretti, P.-Y. 1994. {\it Astron. Astrophys.} 286:701-17.\\
\noindent Kneib, J.-P., Mellier, Y., Pell\'o, R., Miralda-Escud\'e,
J., Le Borgne, J.-F., B\"ohringer, H., Picat, J.-P. 1995. 
 {\it Astron. Astrophys.} 303:27-40.\\
\noindent Kneib, J.-P., Ellis, R. S., Smail, I., Couch, W.,
Sharples, R. M. 1996. {\it Ap. J.} 471:643-56.\\
\noindent Kneib, J.-P., Alloin, D., Mellier, Y., Guilloteau, S., 
Barvainis, R., Antonucci, R. 1998. {\it Astron. Astrophys.} 329:827-39.\\
\noindent Kruse, G., Schneider, P. 1998. Preprint astro-ph/9806071.\\
\noindent Kochanek, C. S., Blandford, R.D., Lawrence, C.R., Narayan, R. 1989. {\it MNRAS} 238:43-56.\\
\noindent Kochanek, C. S. 1990. {\it MNRAS} 247:135-51.\\
\noindent Kochanek, C. S. 1991. {\it Ap. J.} 373:354-68.\\
\noindent Kochanek, C. S. 1995. {\it Ap. J.} 445:559-77.\\
\noindent Kochanek, C. S. 1996. {\it Ap. J.} 466:638-59.\\
\noindent Kovner, I., Milgrom, M. 1987. {\it Ap. J. Lett.} 321:L113-L115.\\
\noindent Krauss, L.M. 1998. Preprint astro-ph/9807376.\\
\noindent Kristian, J., Sachs, R. K. 1966. {\it Ap. J.} 143:379-386.\\
\noindent Kristian, J. 1967. {\it Ap. J.} 147:864-67.\\
\noindent L\'emonon, L., Pierre, M., Cesarsky, C., Elbaz, D., Pell\'o,
R., Soucail, G., Vigroux, L. 1998. {\it Astron. Astrophys. Lett.} 334:L21-L25.\\
\noindent Link, R., Pierce, M. 1998. {\it Ap. J.} 502:63-74.\\
\noindent Lombardi, M., Bertin, G. 1998a. {\it Astron. Astrophys.} 330:791-800.\\
\noindent Lombardi, M., Bertin, G. 1998b. {\it Astron. Astrophys.} 335:1-11.\\
\noindent Lombardi, M., Bertin, G. 1998c. Preprint astro-ph/9806282.\\
\noindent Luppino, G., Kaiser, N. 1997. {\it Ap. J.} 475:20-28.\\
\noindent Lynds, R., Petrosian, V. 1986. {\it BAAS} 18:1014.\\
\noindent Maddox, S.J., Efstathiou, G., Sutherland, W.J., Loveday, J. (1990). {\it
MNRAS} 243:692-712.\\
\noindent Markevitch, M. 1997. {\it Ap. J. Lett.} 483:L17-L20.\\
\noindent  Mart\'{\i}nez-Gonz\'alez, E., Sanz, J. L., Cay\'on, L. 1997.
{\it Ap. J.} 484:1-6.\\
\noindent Mellier, Y. 1997. In {\it The Hubble Space Telescope and the High Redshift
Universe} ed. N.R. Tanvir, A. Arag\'on-Salamnca, J. V. Wall. World Scientific.\\
\noindent Mellier, Y., Fort, B., Soucail, G., Mathez, G., Cailloux, M.
1991. {\it Ap. J.} 380:334-43.\\
\noindent Mellier, Y., Fort, B., Kneib, J.-P. 1993. {\it Ap. J.} 407:33-45.\\
\noindent Mellier, Y., Dantel-Fort, M., Fort, B., Bonnet, H. 1994. {\it Astron. Astrophys.
Lett.} 289:L15-L18.\\
\noindent Mellier, Y., Van Waerbeke, L., Bernardeau, F., Fort, B.
1997. In {\sl Neutrinos, Dark Matter and the Universe}, ed. T. Stolarcyk, J. Tran Thanh Van,
F. Vannucci, pp. 191-204. Fronti\`eres. \\
\noindent Metcalf, B., Silk, J. 1997. {\it Ap. J.} 489:1-6.\\
\noindent Metcalf, B., Silk, J. 1998. {\it Ap. J. Lett.} 492:L1-L4.\\
\noindent Miralda-Escud\'e, J. 1991. {\it Ap. J.} 380:1-8.\\
\noindent Miralda-Escud\'e, J., Babul, A. 1995. {\it Ap. J.} 449:18-27.\\
\noindent Moessner, R., Jain, B. 1998. {\it MNRAS Lett.} 294:L18-L24.\\
\noindent Moessner, R., Jain, B., Villumsen 1998. {\it MNRAS} 294:291-98.\\
\noindent Mould, J., Blandford, R., Villumsen, J., Brainerd, T., Smail,
I., Small, T., Kells, W. 1994. {\it MNRAS} 271:31-38.\\
\noindent Nakamura, T. T. 1997. {\it Publ. Astron. Soc. Japan} 49:151-57.\\
\noindent Nakanishi, K., Ohta. K., Takeuchi, T. T., Akiyama M., Yamada, T.,
Shioya, Y. 1997. {\it Publ. Astron. Soc. Japan} 49:535-38.\\ 
\noindent Navarro, J. F., Frenk, C. S., White, S. D. M. 1997. {\it Ap. J.} 490:
493-508.\\
\noindent Natarajan, P., Kneib, J.-P. 1997. {\it MNRAS} 287:833-47.\\
\noindent Natarajan, P., Kneib, J.-P., Smail, I., Ellis, R. S.
1998. {\it Ap. J.} 499:600-07.\\
\noindent Ota, N., Mitsuda, K., Fukazawa, Y. 1998. {\it Ap. J.} 495:170-78.\\
\noindent Peacock, J.A., Dodds, S. 1996. {\it MNRAS Lett.} 280:L19-L26.\\
\noindent Pell\'o, R., Sanahuja, B., Le Borgne, J.-F., Soucail, G., Mellier, Y. 1991.
{\it Ap. J.} 366:405-11.\\
\noindent Pell\'o, R., Miralles, J.-M., Le Borgne, J.-F., Picat, J.-P.,
Soucail, G., Bruzual, G. 1996. {\it Astron. Astrophys.} 314:73-86.\\
\noindent Pierre, M., Le Borgne, J.-F., Soucail, G., Kneib,
J.-P. 1996. {\it Astron. Astrophys.} 311:413-24.\\
\noindent Puget, J.-L., Abergel, A., Boulanger, F., Bernard, J.-P.,
Burton, W.B., D\'esert, F.-X., Hartmann, D. 1996. {\it Astron. Astrophys.} 308:L5-L8.\\
\noindent R\'efr\'egier, A., Brown, S. T. 1998. Preprint
astro-ph/9803279.\\
\noindent Refsdal, S. 1964. {\it MNRAS} 128:307-310.\\
\noindent Refsdal, S., Surdej, J. 1994. {\it Rep. Prog. Phys.} 56:117-85.\\
\noindent Rodrigues-Williams, L. L., Hogan, C. J. 1994. {\it Astron. J.}  107:451-60.\\
\noindent Rosati, P. 1999. In {\it Wide Field Surveys in Cosmology}, ed. S. Colombi, Y.
Mellier, B. Raban. Fronti\`eres. \\
\noindent Sachs, R.K. 1961. {\it Proc. Roy. Soc. London} A264:309.\\
\noindent Sahu, K. C., Shaw, R., A., Kaiser, M. E., Baum, S. A.,
Fergusson, H. C., Hayes, J. J. E., Gull, T. R., Hill, R. J., Hutchings,
J. B., Kimble, R. A., Plait, P., Woodgate, B. E. 1998. {\it Ap. J.
Lett.} 492:L125-L129.\\
\noindent Sanz, J. L., Mart\'{\i}nez-Gonz\'alez, E., Ben\'{\i}tez, N. 1997. 
 {\it MNRAS} 291:418-24.\\
\noindent Saraniti, D. W., Petrosian, V., Lynds 1996. {\it Ap. J.} 458:57-66.\\
\noindent Schindler, S., Hattori, M., Neumann, D., M., B\"ohringer, H.
1997. {\it Astron. Astrophys.} 317:646-55.\\
\noindent Schneider, P. 1992. {\it Astron. Astrophys.} 254:14-24.\\
\noindent Schneider, P. 1995. {\it Astron. Astrophys.} 302:639-48.\\
\noindent Schneider, P. 1998. {\it Ap. J.} 498:43-47.\\
\noindent Schneider, P., Weiss, A. 1988. {\it Ap. J.} 327:526-43.\\
\noindent Schneider, P., Ehlers, J., Falco, E. E. 1992. {\it
Gravitational Lenses}. Springer.\\
\noindent Schneider, P., Rix, H. W. 1997. {\it Ap. J.} 474:25-36.\\
\noindent Schneider, P., Seitz, C. 1995. {\it Astron. Astrophys.} 294:411-31.\\
\noindent Schneider, P., Kneib, J.-P. 1998. Preprint
astro-ph/9807091. \\
\noindent Schneider, P., Van Waerbeke, L., Mellier, Y., Jain,
B., Seitz, S., Fort, B. 1998a. {\it Astron. Astrophys.} 333:767-78.\\
\noindent Schneider, P., Van Waerbeke, L., Jain, B., Kruse, G. 1998b.
 {\it MNRAS} 296:873-92.\\
\noindent Schramm, T., Kayser, R. 1995. {\it Astron. Astrophys.}
299:1-10.\\
\noindent Seitz, C., Kneib, J.P., Schneider, P., Seitz, S.
1996. {\it Astron. Astrophys.} 314:707-20.\\
\noindent Seitz, C., Schneider, P. 1995a. {\it Astron. Astrophys.} 297:287-99.\\
\noindent Seitz, C., Schneider, P. 1997. {\it Astron. Astrophys.} 318:687-99.\\
\noindent Seitz, S. 1999. In {\it Wide Field Surveys in Cosmology}, ed. S. Colombi, Y.
Mellier, B. Raban. Fronti\`eres. \\
\noindent Seitz, S., Schneider, P. 1995b. {\it Astron. Astrophys.} 302:9-20.\\
\noindent Seitz, S., Schneider, P.  1996. {\it Astron. Astrophys.} 305:383-401.\\
\noindent Seitz, S., Saglia, R. P., Bender, R., Hopp, U.,
Belloni, P., Ziegler, B. 1997. Preprint astro-ph/9706023.\\
\noindent Seitz, S., Schneider, P., Bartelmann, M. 1998. 
{\it Astron. Astrophys.} 337:325-37.\\
\noindent Seljak, U. 1996. {\it Ap. J.} 463:1-7.\\
\noindent Seljak, U. 1997a. Preprint astro-ph/9711124.\\
\noindent Seljak, U. 1997b. {\it Ap. J.} 482:6-16.\\
\noindent Smail, I. (1993). {\it Gravitational Lensing by Rich Clusters of Galaxies}.
PhD Thesis.\\
\noindent Smail, I., Ellis, R. S., Arag\'on-Salamanca, A., Soucail, G., 
Mellier, Y., Giraud, E. 1993. {\it MNRAS} 263:628-40.\\
\noindent Smail, I., Ellis, R. S., Fitchett M. 1994. {\it MNRAS} 270:245-70.\\
\noindent Smail, I., Ellis, R. S., Fitchett M., Edge, A. C. 1995. 
{\it MNRAS} 273:277-94.\\
\noindent Smail, I., Couch, W., Ellis, R. S., Sharples, R. M. 1995.
{\it Ap. J.} 440:501-09.\\
\noindent Smail, I., Dickinson, M. 1995. {\it MNRAS} 455:L99-L102.\\
\noindent Smail, I., Hogg, D.W., yan, L., Cohen, J.G. 1995. {\it Ap. J. Lett.} 
449:L105-L108.\\
\noindent Smail, I., Ivison, R. J., Blain, A. W. 1997. 
{\it Ap. J. Lett.} 490:L5-L8.\\
\noindent Smail I., Ellis, R. S., Dressler, A., Couch, W. J., Oemler, A.,
Sharples, R., Butcher, H. 1997. {\it Ap. J.} 479:70-81.\\
\noindent Smail, I., Ivison, R. J., Blain, A. W., Kneib, J.-P. 1998.
Preprint astro-ph/9806061.\\
\noindent Soifer, B.T., Neugebauer, G., Franx, M., Matthews, K., Illingworth, G.D.
1998. Preprint astro-ph/9805219.\\
\noindent Soucail, G., Fort, B., Mellier, Y., Picat, J.-P. 1987. 
{\it Astron. Astrophys. Lett.} 172:L14-L17.\\
\noindent  Soucail, G.,  Mellier, Y., Fort, B., Mathez, G.,
Cailloux, M. 1998. {\it Astron. Astrophys. Lett.} 191:L19-L22.\\
\noindent Squires, G., Kaiser, N., Babul, A., Fahlman, G.,
Woods, D., Neumann, D.M., B\"ohringer, H. 1996a. {\it Ap. J.} 461:572-86.\\
\noindent Squires, G., Kaiser, N., Fahlman, G., Babul, A., Woods,
D. 1996b. {\it Ap. J.} 469:73-77.\\
\noindent Squires, G., Kaiser, N. 1996. {\it Ap. J.} 473:65-80.\\
\noindent Squires, G., Neumann, D.M., Kaiser, N., Arnaud, M.,
Babul, A., B\"ohringer, H. Fahlman, G., Woods, D. 1997. {\it Ap. J.} 482:648-58.\\
\noindent Stebbins, A. 1996. Preprint astro-ph/9609149.\\
\noindent Stebbins, A. 1999.   In {\it Wide Field Surveys in Cosmology}, ed. S.
Colombi, Y.  Mellier, B. Raban. Fronti\`eres. \\
\noindent Steidel, C., Adelberger, K., Giavalisco, M., Dickinson, M., 
Pettini, M., Kellogg, M. 1998, in {\it The Young Universe}, ed. S.
D'Odorico, A. Fontana, E. Giallongo, pp. 428-35. PASP Conf. Series. Vol. 416.
\\ 
\noindent Stompor, R., Efstathiou, G. 1998. Preprint astro-ph/9805294.\\
\noindent Szapudi, I., Colombi, S. 1996. {\it Ap. J.} 470:131-48.\\
\noindent Taylor, A. N., Dye, S., Broadhurst, T. J., Ben\'{\i}tez, N.,
Van Kampen, E. 1998. {\it Ap. J.} 501:539.\\
\noindent Tomita, K., Watanabe, K. 1990. {\it Prog. Theo. Phys.} 83:467-90.\\
\noindent Trager, S. C., Faber, S. M., Dressler, A., Oemler, A. 1997. 
 {\it Ap. J.} 485:92-99.\\
\noindent Tyson, J.A. 1985. {\it Nature} 316:799-800.\\
\noindent Tyson, J.A., Valdes, F., Jarvis, J. F., Mills, A. P. 1984. 
 {\it Ap. J. Lett.}  281:L59-L62.\\
\noindent Tyson, J.A. 1988. {\it Astron. J.} 96:1-23.\\
\noindent Tyson, J.A., Valdes, F., Wenk, R. A. 1990. {\it Ap. J. Lett.}
349:L1-L4.\\
\noindent Tyson, J.A., Fischer, P. 1995. {\it Ap. J. Lett.} 446:L55-L58.\\
\noindent Tyson, J.A., Kochanski, G. P., Dell'Antonio I. P. 1998. 
 {\it Ap. J.  Lett.} 498:L107-L110.\\
\noindent Valdes, F., Tyson, J. A., Jarvis, J. F. 1983. {\it Ap. J.} 271:431-41.\\
\noindent Van Kampen, E. 1998. Preprint astro-ph/9807035.\\
\noindent Van Waerbeke, L. 1998a. {\it Astron. Astrophys.} 334:1-10.\\
\noindent Van Waerbeke, L. 1998b. Preprint astro-ph/9807041.\\
\noindent Van Waerbeke, L. 1999. Submitted to {\it MNRAS}.\\
\noindent Van Waerbeke, L., Mellier, Y. 1997. in {\it proceedings
of the XXXIst Rencontres de Moriond}. Fronti\`eres.\\
\noindent Van Waerbeke, L., Mellier, Y., Schneider, P., Fort, B., Mathez, G.
1997. {\it Astron. Astrophys.} 317:303-17.\\
\noindent Van Waerbeke, L., Bernardeau, F., Mellier, Y. 1998. Preprint
astro-ph/9807007.\\
\noindent Villumsen, J. V. 1996. {\it MNRAS} 281:369-83.\\
\noindent Wambsganss, J., Cen, R., Ostriker, J. P. 1998. {\it Ap. J.} 494:29-46.\\
\noindent Wallington, S., Kochanek, C. S., Koo, D. C. 1995. 
{\it Ap. J.} 441:58-69.\\
\noindent Walsh, D., Carswell, R.F., Weymann, R.J. 1979. {\it Nature} 279:381.\\
\noindent Webster, R. 1985. {\it MNRAS} 213:871-88.\\
\noindent White, M. 1998. Preprint astro-ph/9802295.\\
\noindent White, M., Scott, D., Silk, J. 1994. {\it Annu. Rev. Astron.
Astrophys.} 32:319-70.\\
\noindent Williams, L. L. R., Irwin, M. 1998. 
{\it MNRAS} 298:378-86.\\
\noindent Wilson, G., Cole, S., Frenk, C. S. 1996a. {\it MNRAS} 280:199-218.\\
\noindent Wilson, G., Cole, S., Frenk, C. S. 1996b. {\it MNRAS} 282:501-10.\\
\noindent Wu, X.-P., Mao, S. 1996. {\it ApJ} 463:404-08.\\
\noindent Wu, X.-P., Fang, L.-Z. 1996. {\it Ap. J.} 461:L5-L8.\\
\noindent Wu, X.-P., Fang, L.-Z. 1997. {\it Ap. J.} 483:62-67.\\
\noindent Zaldarriaga, M., Seljak, U. 1998. Preprint astro-ph/9803150.\\
\noindent Zel'dovich, Y. B. 1964. {\it Sov. Astron.} 8:13-16.\\
\noindent Zhu, Z.-H., Wu, X.-P. 1997. {\it Astron. Astrophys.} 324:483.\\
\noindent Zwicky, F. 1933. {\it Helv. Phys. Acta} 6:10.\\
\noindent Zwicky, F. 1937. {\it Phys. Rev.} 51:290.\\
}
\end{document}